%% file: ms.tex
\documentclass[]{emulateapj}

\usepackage{graphics}
\usepackage{mathptmx}
\usepackage{natbib}

\newcommand{\teff}{$T_{\!\mbox{\scriptsize\em eff}}$}

\newcommand{\msun}{$M_\odot$}

\newcommand{\hi}{H\,{\sc i}\rm}
\newcommand{\hei}{He\,{\sc i}\rm}

\newcommand{\oiii}{[O\,{\sc iii}]}

 \newcommand{\hii}{\ion{H}{2}}

\slugcomment{To appear in Astrophysical Journal}

\shorttitle{Metallicity and Distance of NGC\,3621}
\shortauthors{Kudritzki et al.}

\begin{document}


\title{Stellar Metallicity of the Extended Disk and Distance of the Spiral Galaxy NGC\,3621}



\author{Rolf-Peter Kudritzki\altaffilmark{1}}
\affil{Institute for Astronomy, University of Hawaii, 2680 Woodlawn Drive, Honolulu, HI 96822, USA}
\email{kud@ifa.hawaii.edu}
\author{Miguel A. Urbaneja}
\affil{Institute for Astro- and Particle Physics, University of Innsbruck, Technikerstr. 25/8, A-6020 Innsbruck, Austria}
\email{Miguel.Urbaneja-Perez@uibk.ac.at}
\author{Fabio Bresolin}
\affil{Institute for Astronomy, University of Hawaii, 2680 Woodlawn Drive, Honolulu, HI 96822, USA}
\email{bresolin@ifa.hawaii.edu}
\author{Matthew W. Hosek Jr.}
\affil{Institute for Astronomy, University of Hawaii, 2680 Woodlawn Drive, Honolulu, HI 96822, USA}
\email{mwhosek@ifa.hawaii.edu}

\and

\author{Norbert Przybilla}
\affil{Institute for Astro- and Particle Physics, University of Innsbruck, Technikerstr. 25/8, A-6020 Innsbruck, Austria}
\email{Norbert.Przybilla@uibk.ac.at}\email{Miguel.Urbaneja-Perez@uibk.ac.at}




\begin{abstract}
  Low resolution ($\sim 4.5$ \AA) ESO VLT/FORS spectra of blue supergiant stars are analyzed
  to determine stellar metallicities (based on elements such as iron, titanium, magnesium)
  in the extended disk of the spiral galaxy NGC\,3621. Mildly subsolar metallicity 
  (-0.30 dex) is found for the outer objects beyond 7 kpc independent of galactocentric 
  radius and compatible with the absence of a metallicity gradient confirming the results 
  of a recent investigation of interstellar medium \hii~region gas oxygen abundances. 
  The stellar metallicities are slightly higher than those from the \hii~regions when based 
  on measurements of the weak forbidden auroral oxygen line at 4363~\AA~but lower than the ones obtained 
  with the R$_{23}$ strong line method. It is shown that the present level of metallicity 
  in the extended disk cannot be the result of chemical evolution over the age of the disk with the present 
  rate of in situ star formation. Additional mechanisms must be involved. In addition to metallicity, 
  stellar effective temperatures, gravities, interstellar reddening, and bolometric magnitudes are 
  determined. After application of individual reddening corrections for each target the
  flux-weighted gravity-luminosity relationship of blue supergiant stars is used to obtain
  a distance modulus of 29.07 $\pm$ 0.09 mag (distance $D=6.52\pm0.28$\,Mpc). This new distance is discussed 
  in relation to Cepheid and tip of the red giant branch distances.
\end{abstract}



\altaffiltext{1}{University Observatory Munich, Scheinerstr. 1, D-81679 Munich, Germany}

\section{Introduction}

NGC\,3621 is a bulgeless isolated galaxy with a relatively regular spiral structure which 
extends out to at least two isophotal radii. The galaxy with its extended disk is imbedded into 
an envelope of neutral hydrogen gas (\citealt{koribalski04}). A recent study by \citet{bresolin12} 
analyzing the emission lines of \hii~region spectra reveals that the extended disk (beyond 0.8 
isophotal radii) has no oxygen abundance gradient and shows an oxygen abundance distribution which
is spatially flat at a relatively high level of -0.4 dex below solar oxygen abundance. While the flatness, 
though striking, does not seem to be a major issue and according to Bresolin et al. can be 
explained by the fact that the star formation efficiency is almost constant with radius in the 
extended disk, it is the high level of metallicity which poses a problem. Bresolin et al. estimate 
that with the large neutral hydrogen gas reservoir and the low-level of ongoing star formation the 
time required to enrich the interstellar gas to the observed level is 10~Gyr. While inner disks may 
have such an age, Bresolin et al. argue that outer disks are at least a factor of two younger,
Thus, chemical enrichment of the interstellar medium through in situ star formation at the present 
level could have resulted only in a very low oxygen abundance with values smaller than -0.7 dex below the solar 
abundance. As possible mechanisms to explain the discrepancy Bresolin et al. discuss radial metal transport 
from the inner to the outer disk and gas accretion from the intergalactic medium by metal enriched galactic 
outflows.

There are many galaxies with extended star forming disks and a flat oxygen abundance distribution, 
for instance M83 (\citealt{bresolin09b}), NGC\,4624 (\citealt{goddard11}), NGC\,1512 (\citealt{bresolin12}) 
and the thirteen mostly merging galaxies investigated by \citet{werk11}. However, NGC\,3621  
appears to be the poster example for a galaxy with a well defined inner abundance gradient and a 
large outer extended disk with a flat metallicity distribution (see Figure 9 of \citealt{bresolin12}).
In addition, NGC\,3621 is also the case where the discrepancy between expected and observed outer 
disk metallicity, as discussed above, is most significant. We have, thus, decided to reinvestigate 
the metallicity distribution in the disk of this galaxy using an independent alternative method, the 
spectral analysis of blue supergiant stars (BSGs). With absolute magnitudes  up to  $M_{V} \cong -10$
BSGs are the brightest star in the universe at visual light and perfect objects for quantitative 
stellar abundance studies beyond the Local Group. They are massive stars in the range between 15 
and 40~\msun. At an age of $\sim$ 10 million years they have left the main sequence and cross the 
Hertzpsrung-Russell diagram to become red supergiant stars and then to explode as core collapse 
supernovae. \citet{kud08} in their study of metallicity in the Sculptor spiral galaxy NGC\,300 were the
first to demonstrate how accurate metallicities based on elements such as iron, chromium,
titanium etc. can be determined from low resolution spectra of individual BSGs using model atmosphere 
techniques. Since then a large number of galaxies has been studied (WLM -- \citealt{bresolin06}; 
\citealt{urbaneja08}; NGC\,3109 -- \citealt{evans07}, \citealt{hosek14}; IC\,1613 --
\citealt{bresolin07}; M33 -- \citealt{u09}; NGC\,55 -- \citealt{castro12}; M81 -- \citealt{kud12}; 
NGC\,4258 -- \citealt{kud13}).

\begin{figure}[!]
 \begin{center}
  \includegraphics[width=0.45\textwidth]{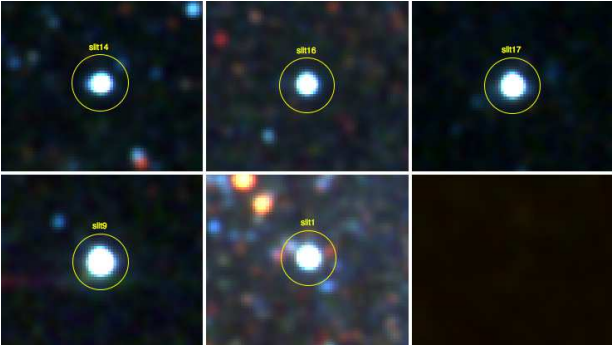}
  \caption[]{
Enlarged B, V, I composite HST ACS images of five of the observed BSG targets. The circle corresponds to 1 arcsec diameter. \label{targets} }
 \end{center}
\end{figure}

A particular motivation for this work is the systematic uncertainty inherent in \hii~region abundance 
studies. In most cases they are based on the use of the strongest nebular emission lines and using 
oxygen as a proxy for stellar metallicity. However, as shown, for instance, by 
\citet{kud08}, \citet{kewley08}, \citet{bresolin09a}, \citet{u09}, \citet{kud12} these 
``strong-line methods'' are subject to systematic uncertainties as large as 0.6 dex. They are poorly 
understood and can severely affect the values of galaxy metallicities and abundance gradients.
\citet{bresolin12} were aware of this uncertainty in their investigation of NGC\,3621 and used three
different strong line calibrations. They all resulted in a flat oxygen abundance distribution over 
the outer extended disk, however the abundance level was different. Two calibrations yielded -0.44 dex 
lower than solar, whereas the third showed a much higher value, only -0.08 dex below and, thus, almost 
solar. In addition to the calibration dependent use of strong lines \citet{bresolin12} were also able to 
determine \hii~region electron temperatures from the detection of the weak auroral [OIII] 4363 line, 
which was detected in 12 of the observed 72 \hii~regions. Nebular oxygen abundance determinations based 
on the temperature information of this line are more reliable, as it can be used to calulate the 
excitation of the upper levels of the strong lines and thus their emission coefficient. For instance, 
in the case of the galaxy NGC\,300 excellent agreement has been found by \citet{bresolin09a} between 
\hii~region and BSG metallicities. On the other hand, the work by \citet{stasinska05}, \citet{bresolin05}, 
\citet{ercolano10}, and \citet{zurita12} indicates that also this method might be subject to 
uncertainties albeit much smaller than the strong line methods. An additional, more general problem of 
the oxygen abundances obtained from \hii~regions might be the depletion of oxygen through the formation 
of dust grains (for a discussion see \citealt{bresolin09a}). While the BSG abundance determination may 
certainly also be subject to systematic uncertainties, we see a double advantage in the investigation of BSG 
metallicties in the disk of NGC\,3621. First, it simply provides an independent measurement of 
metallicity using a well established accurate method. Second, it will give information about metallicity 
from elements other than oxygen and more relevant to ``metallicity'' in the sense of chemical evolution.
We hope to use this advantage to develop improved constraints on the metallicity enrichment of the 
outer disks of star forming galaxies.                

\begin{figure}
\begin{center}
  \includegraphics[scale=0.35,angle=90]{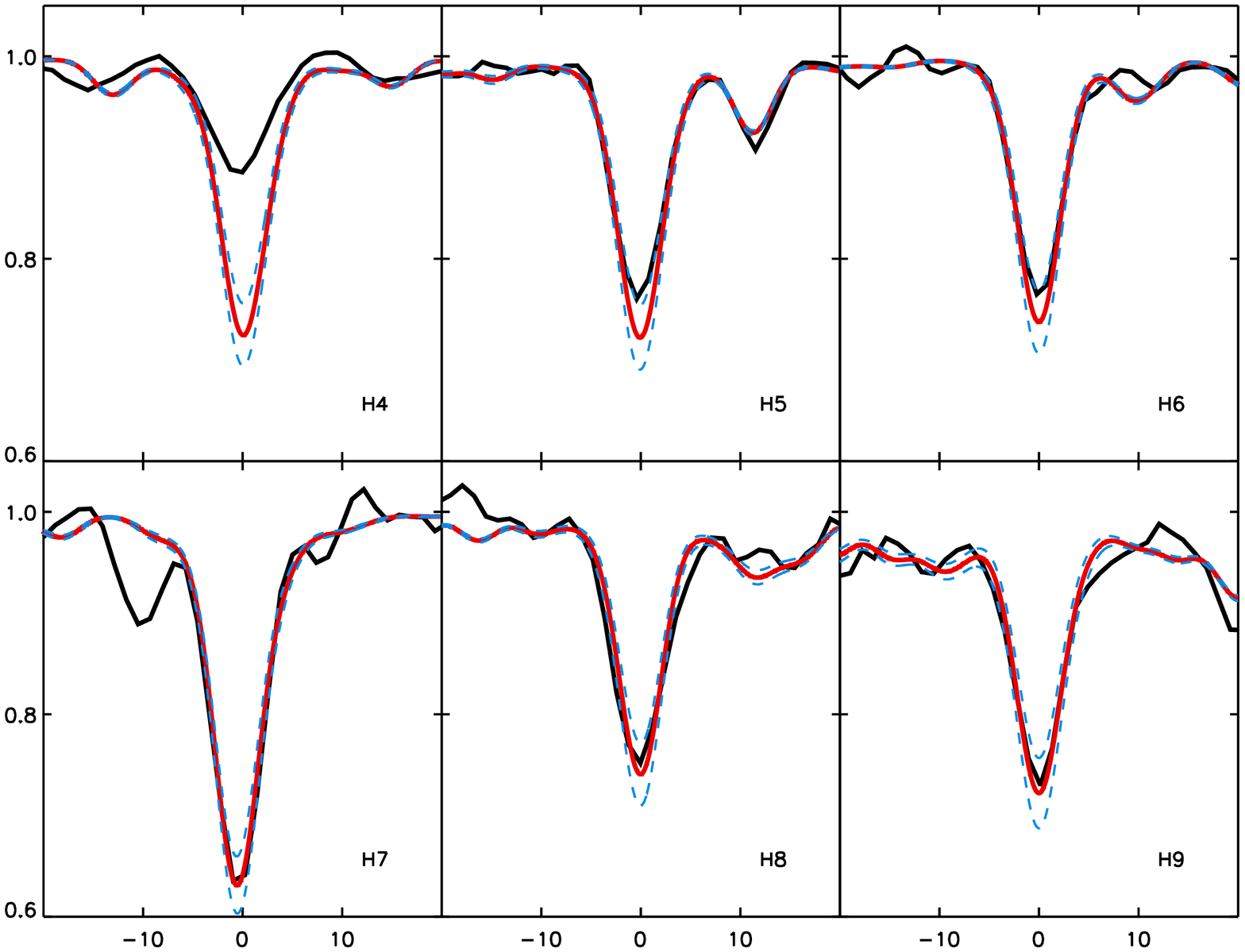}
  \includegraphics[scale=0.35,angle=90]{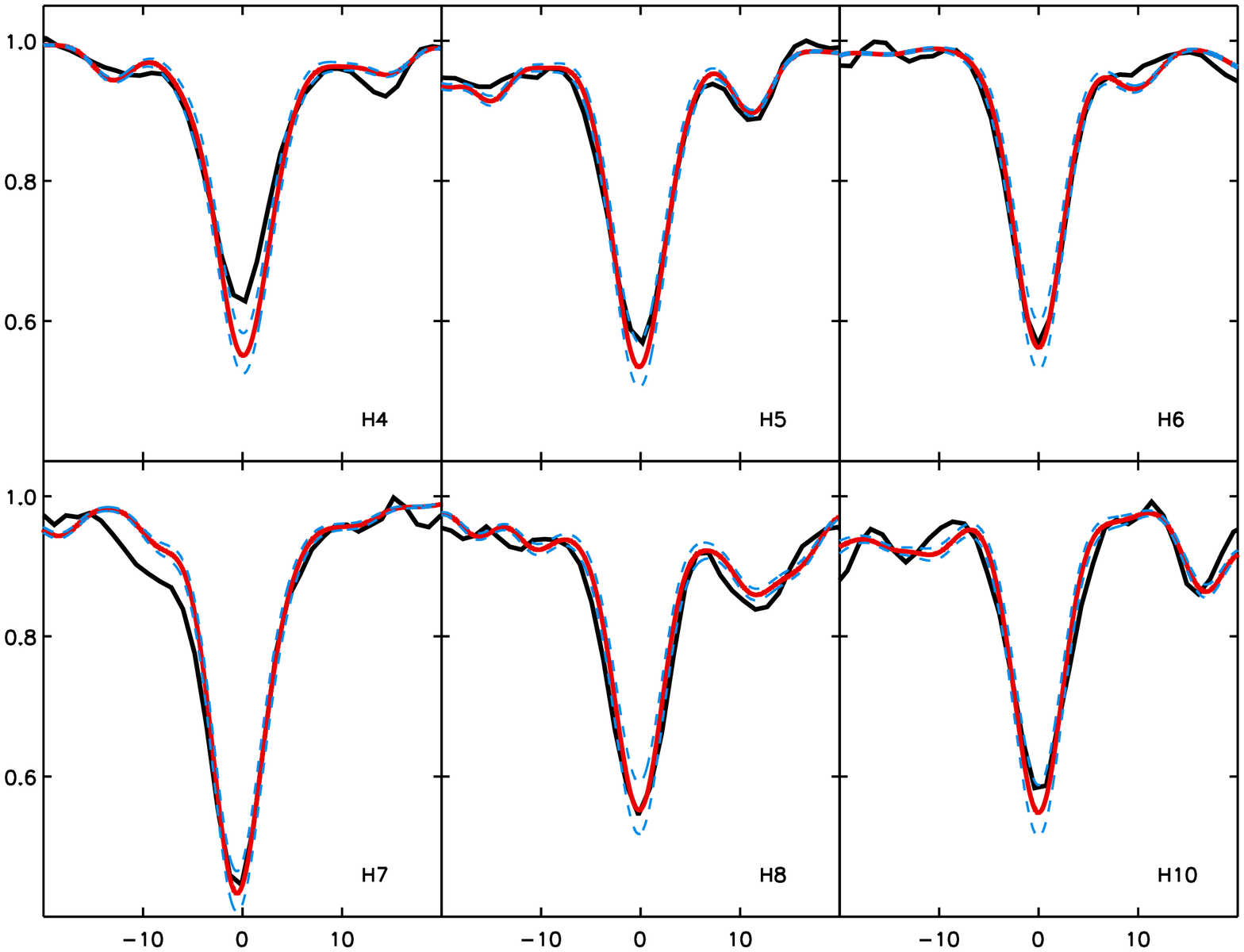} 
\caption{Top: Fit of observed Balmer line profiles (black solid) of target Slit 9 with model atmospheres of \teff=8750K and 
log g = 1.00 (red, solid) and 0.95 and 1.05 (both blue dashed), respectively. Bottom: Similar fit of target Slit 17 with 
\teff=8100K and log g =0.90 (red, solid) and 0.85, 0.95 (blue dashed). The gravities log g are given in cgs units.  \label{balmfit_s9_s17}}
 \end{center}
\end{figure}

NGC\,3621 is not only interesting with regard to disk evolution and star formation. It is also an 
important galaxy for the determination of extragalactic distance scale. With an inclination angle of 65 
degrees \citep{deblok08} and well ordered \hi~rotation it is an ideal galaxy for the calibration of the 
Tully-Fisher method. It is, thus, not surprising that it has been included in the Hubble Space Telescope 
Extragalactic Distance Scale Key Project (\citealt{freedman01}). Also the HST tip of the red giant branch 
(TRGB) study to calibrate type I supernovae as distance indicators by \citet{mould08} has included this 
important galaxy. Distance moduli determined over the last 13 years vary between 28.9 and 29.4 mag 
(see NED database http://ned.ipac.caltech.edu). Our BSG spectroscopic study provides an important alternative for 
distance determination through the use of the flux-weighted gravity -- luminosity relationship (FGLR). 
This technique, which uses the determination of stellar temperature and gravities to predict absolute 
bolometric magnitudes, has been introduced by \citet{kud03} and \citet{kud08} and has already been 
applied successfully in a variety of cases (WLM -- \citealt{urbaneja08}; NGC\,3109 -- \citealt{hosek14}; 
M33 -- \citealt{u09}; M81 -- \citealt{kud12}). 

\begin{figure}
\begin{center}
 \includegraphics[scale=0.35,angle=90]{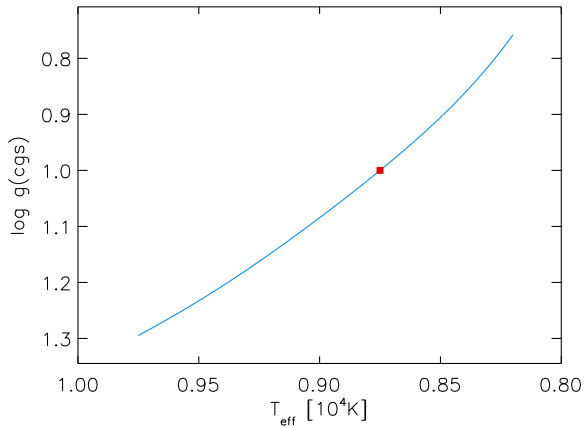}
\caption{Fit curve in the gravity-temperature plane for target Slit 9 along which the calculated Balmer line profiles agree with the observations. 
The fit point of Figure 2 (top) is indicated by the red square.   \label{balmiso_s9}}
 \end{center}
\end{figure}

\section{Target Stars and Observations}

The observations of the BSG target stars were carried out in 2000 March 1-2 with the focal reducer low-dispersion 
spectrograph FORS1 attached to the ESO VLT and are described in detail in Bresolin, Kudritzki, Mendez 
\& Pryzibilla (2001, hereinafter BKMP). In summary, several exposures of 
one single multislit setup were taken with a total integration time of 10.7 hours using a 600 groove/mm grism 
and slitlets with a width of 1 arcsecond. The spectra cover a range of 3700 to 5000~\AA. The nominal resolution 
is 5~\AA, however, with 0.8 arcsec seeing conditions the effective resolution obtained was slightly higher. For the 
quantitative spectral analysis described in the next section the spectra were box smoothed over three pixel 
(1 pixel corresponds to 1.195~\AA). After smoothing the spectral resolution measured was about 4.5~\AA. The 
signal-to-noise ratio of the smoothed spectra varies between 120 and 45 depending on the brightness of the objects 
and how well they were centered into the slitlets.

\input{tab1}

The observations were part of the FORS Guaranteed Time Observing (GTO) program and were thought as a pilot 
study to demonstrate the feasibility of spectroscopy of individual supergiant stars in galaxies 
at distances significantly beyond the Local Group. When the observed spectra were presented in BKMP, 
the method for a consistent quantitative analysis based on model atmospheres of the low resolution spectra still 
had to be developed. As a first step, BKMP determined spectral types. For one supergiant target of 
spectral type A (Slit 9) they used the spectral type to estimate effective temperature and calculated model 
atmospheres at this temperature to qualitatively estimate metallicity. LMC metallicity was found for this target.

\begin{figure}
\begin{center}
 \includegraphics[scale=0.35,angle=90]{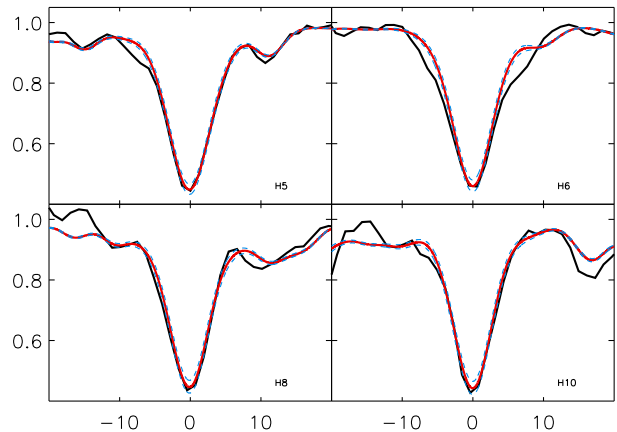}
\caption{Fit of observed Balmer line profiles (black solid) of target Slit 14 with model atmospheres of \teff=8300K and 
log g = 1.25 (red, solid) and 1.20 and 1.30 (both blue dashed), respectively. \label{balmfit_s14}}
 \end{center}
\end{figure}

\begin{figure}
\begin{center}
 \includegraphics[scale=0.5]{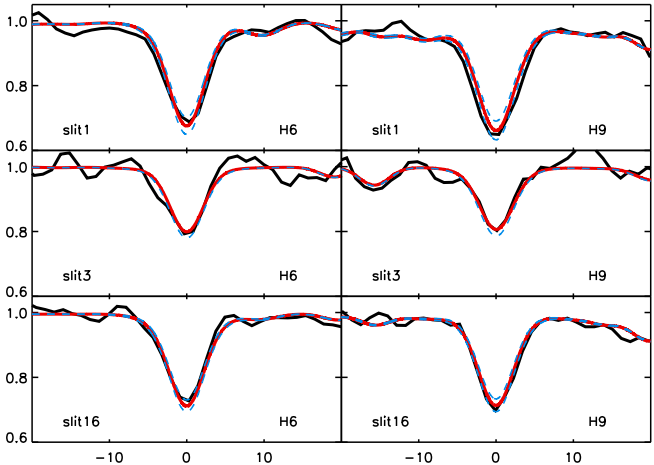}
\caption{Fit of observed Balmer line profiles H$_{6}$ and H$_{9}$ of targets Slit 1, 3, and 16 (black solid). The model atmopsheres (red, solid) 
used are \teff = 9000K, log g = 1.20 for Slit 1, \teff = 13500K, log g = 1.70 for Slit 3 and \teff = 10500K, log g = 1.50 for Slit 16. Additional 
models with log g 0.05 dex smaller or larger are plotted as blue-dashed.  \label{balmfit_smulti}}
 \end{center}
\end{figure}

However, \citet{evans03} correctly pointed out that because the relationship between spectral type and effective 
temperature is metallicity dependent the uncertainties introduced in this way can be significant. Thus, a new method 
was needed providing a self-consistent determination of temperatures, gravities and metallicities. Such a method was 
developed by \citet{kud08} in their work on BSGs in the Sculptor galaxy NGC 300 and further improved in the 
subsequent papers refered to in the previous section. The application of this new method on the old FORS spectra 
of the year 2000 enables us now to resume the quantititave spectroscopic study originally intended.

18 stellar targets were originally selected by BKMP from broad-band color magnitude diagrams (B, V, I CCD frames 
taken with VLT/FORS) after careful inspection of the ground-based images (FWHM $\approx$ 0.6 arcsec) with respect 
to blending with other sources. They were then observed with the single multislit setup. Of these 18 targets eight showed spectra 
of isolated individual BSGs. The other objects were either emission line stars with spectra resembling luminous 
blue variables (LBV) or displayed composite spectra or very strong ISM nebular emission contaminating the stellar 
absorption lines. Seven of the BSGs had the right spectral type A or B for which our model atmosphere analysis 
method is applicable. Unfortunately, for one of these BSGs the signal to noise ratio of the observed spectrum 
turned out to be insufficient for a spectral analyis. Thus, we were left with six BSG objects for our spectroscopic 
study. The observed properties of these objects are given in Table 1. Their location within NGC\,3621 is shown in 
Figure 1 of BKMP.

\begin{figure}
\begin{center}
 \includegraphics[scale=0.35,angle=90]{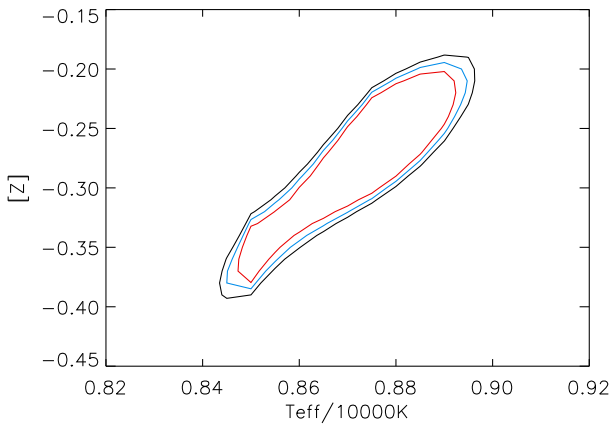}
\caption{Isocontours $\Delta \chi^{2}$ for target Slit 9 in the metallicity [Z] and effective temperature \teff plane obtained from the fit of metal lines. 
$\Delta \chi^{2}$=3 (red), 6 (blue), 9 (black), respectively, are plotted. \label{iso_s9}}
 \end{center}
\end{figure}

NGC\,3621 has also been studied with HST/ACS (HST-GO-9492, P.I.: Fabio Bresolin) in three  filters, F435W, F555W, 
and F814W. Two ACS fields are available which include five of our six targets. PSF-photometry of the stellar sources was carried out of 
all the individual uncombined bias subtracted, flat-fielded images available from the STScI archive. The ACS module included in the DOLPHOT
software package (\citealt{dolphin02}) was used together with a drizzle image for each pointing as a reference frame corresponding to the F555W filter.
ACS/DOLPHOT can be applied with a variation of a large set of parameters. Here, the parameter values of the ANGST survey \citep{dalcanton09} 
were replicated and used. The magnitudes obtained in the ACS filter system were then transformed to the Johnson-Cousins system.
B, V, I magnitudes are given in Table 1. Object Slit 3 of Table 1 is not covered by the HST observations 
and only ground based BVI photometry published by BKMP is available. On the other hand, for ten targets of BKMP listed 
their Table 1 we were able to obtain HST/ACS photometry allowing us to apply a transformation of BKMP to HST/ACS 
photometry for Slit 3. With magnitude differences between HST and FORS $\Delta$m = -0.013, -0.097 and -0.126 mag for 
B, V, and I, respectively, we obtain the photometric values for Slit 3 in Table 1.

\begin{figure}
\begin{center}
 \includegraphics[scale=0.3,angle=90]{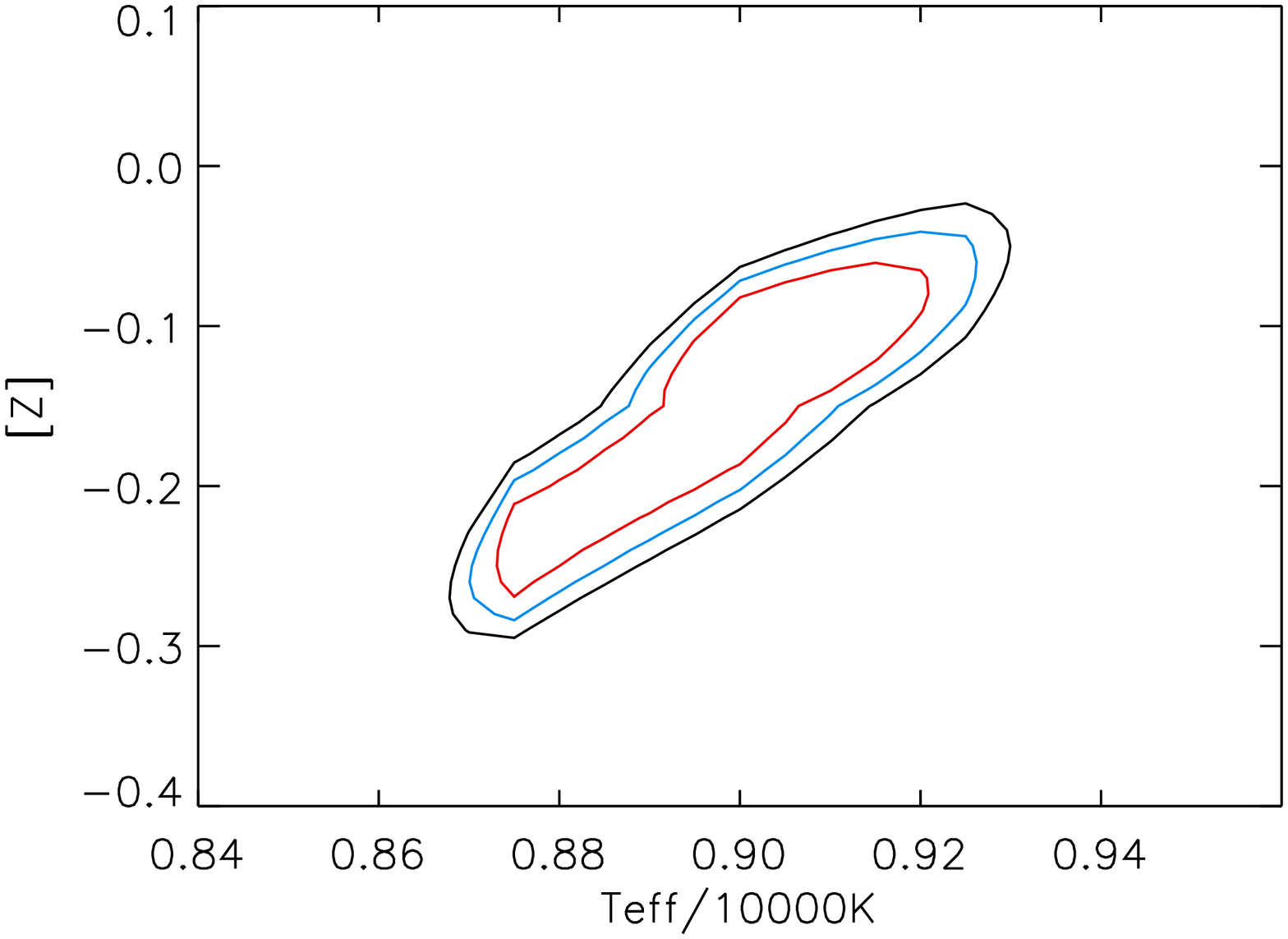}
 \includegraphics[scale=0.3,angle=90]{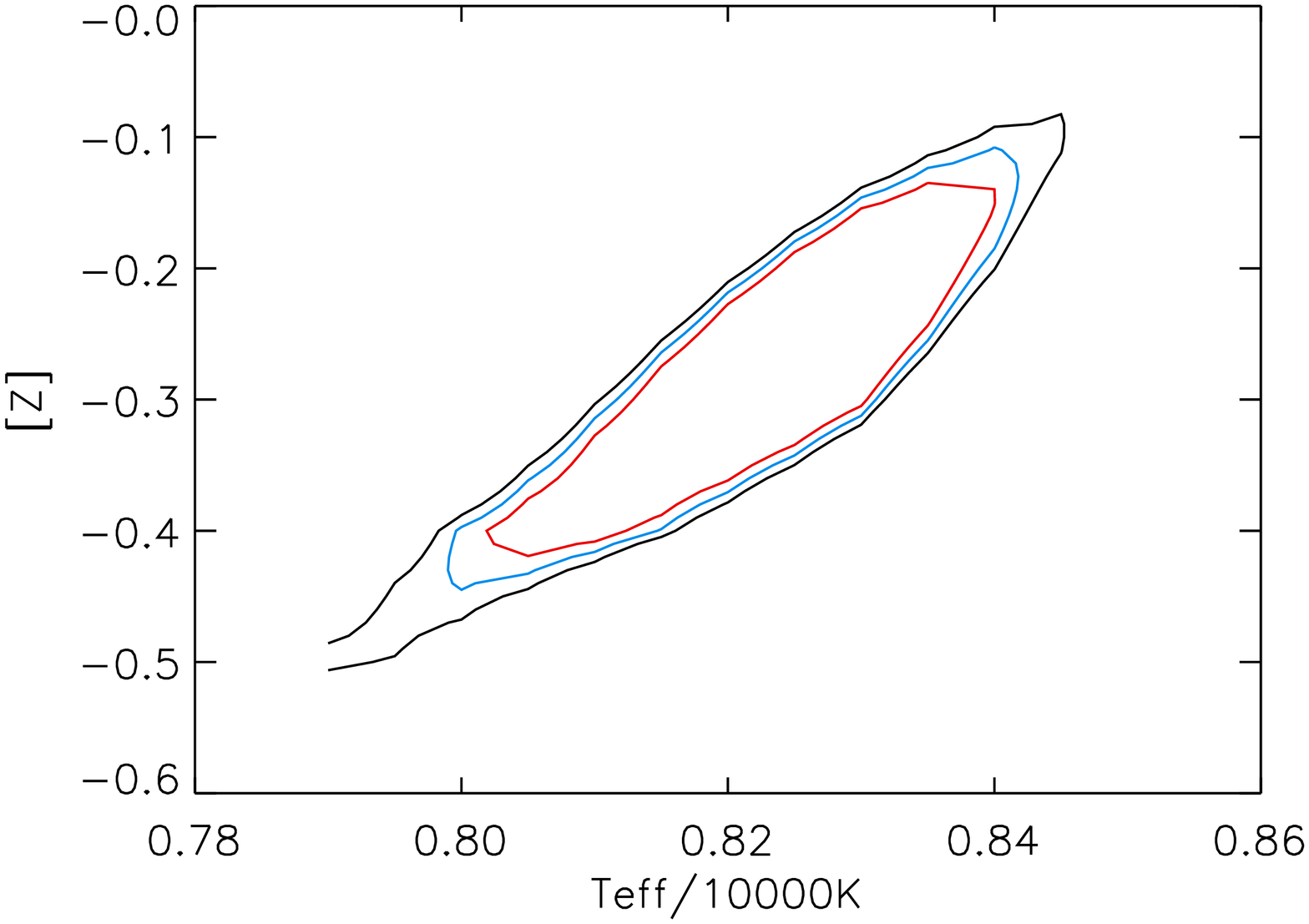}
 \includegraphics[scale=0.3,angle=90]{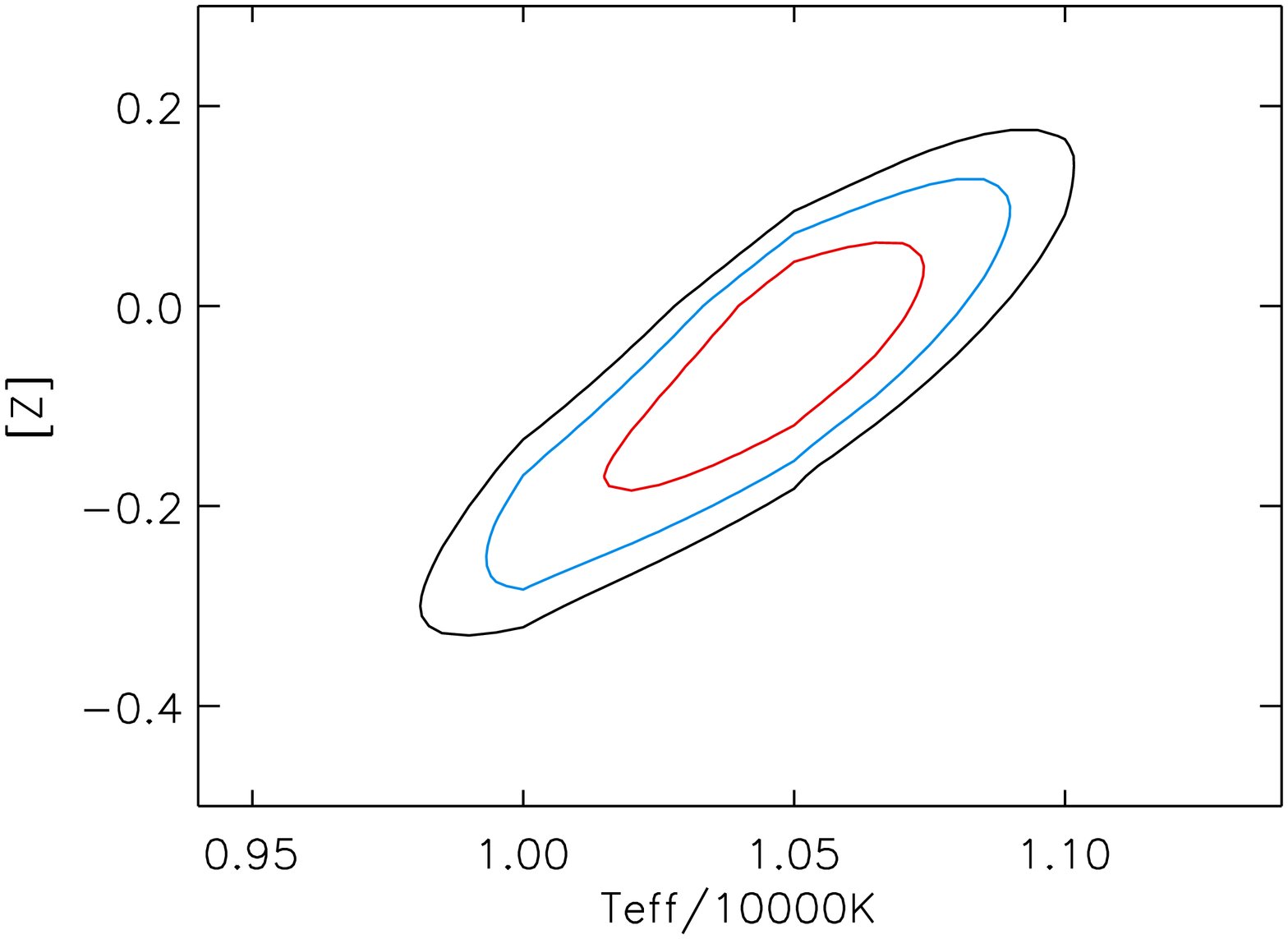}
 \includegraphics[scale=0.3,angle=90]{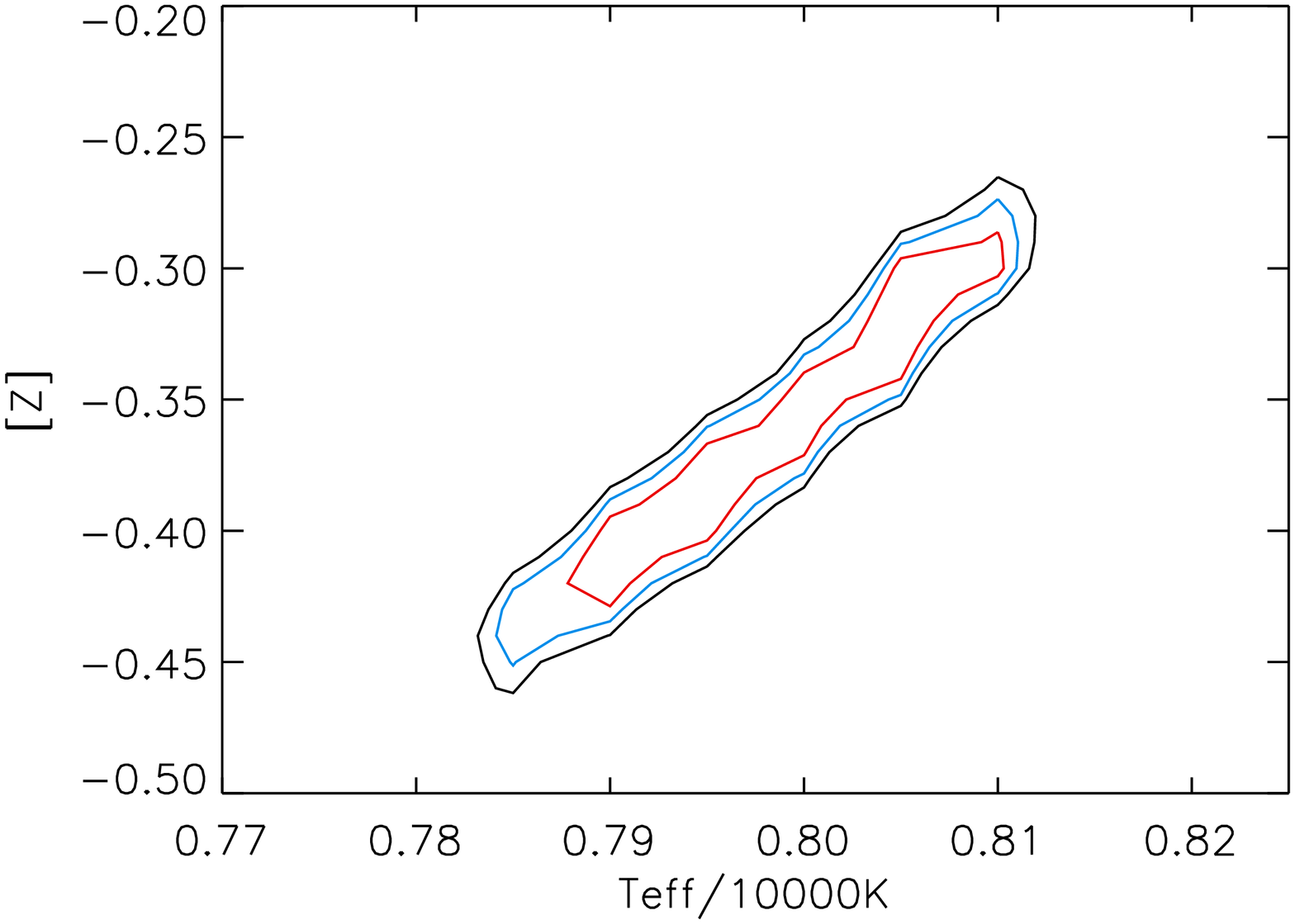}
\caption{Isocontours $\Delta \chi^{2}$ for targets Slit 1 (top left), Slit 14 (top right), Slit 16 (bottom left), Slit 17 (bottom right).
$\Delta \chi^{2}$=3 (red), 6 (blue), 9 (black). \label{iso_multi}}
 \end{center}
\end{figure}

Figure~\ref{targets} shows enlarged B, V, I composite images of the five targets observed with HST/ACS and demonstrate that
the selection based on high quality FORS images worked well. The targets are fairly isolated. For each of them the 
brightest neighbor object within a box of one arcsec diameter is at least three magnitudes fainter.

In their distribution of galactocentric distance the BSGs studied cover the range from 0.6 to 1.3 in R/R$_{25}$ (R$_{25}$
is the isophotal radius, see Table 1). According to \citet{bresolin12} (see their Figure 9) this is the range in 
galactocentric distance where the metallicity gradient of the inner disk flattens and the metallicity distribution 
becomes constant (see also figure~\ref{metal}). At the same time, the targets 
cover a range of one magnitude in brightness sufficient for an application of the FGLR-method to determine a distance.
   
\section{Spectroscopic Analysis}

The goal of the spectroscopic analysis is the determination of stellar effective temperature, gravity and metallicity 
for our BSG targets. For this purpose we compare the normalized observed spectra with synthetic spectra of a comprehensive 
grid of line-blanketed model atmospheres and detailed NLTE line formation calculations. The grid of model spectra and 
the analyis method are described in Kudritzki et al. (2008, 2012, 2013) and \citet{hosek14}. The analyis progresses 
in several steps. The first step is the fit of the Balmer lines. We estimate an effective temperature \teff~from the 
spectral type and determine the gravity log g at which we obtain the best fit of the Balmer lines. An example is given 
in Figure~\ref{balmfit_s9_s17} for targets Slit 9 and 17. The comparison of the different higher Balmer lines H$_{6,7,8,9,10}$   
indicates that a fit of log g accurate to 0.05 dex or better is possible at a fixed temperature \teff. H$_{4}$ is often 
contaminated by stellar wind and \hii~region emission. The effect is clearly present in the examples of 
figure~\ref{balmfit_s9_s17}. In cases of very strong stellar winds and \hii~emission this can 
also affect H$_{5}$ (see figure~\ref{balmfit_s9_s17})  and H$_{6}$ (Slit 9 in figure~\ref{balmfit_s9_s17}) .  
H$_{7}$ can sometimes be blended by interstellar CaII depending on the strength of 
interstellar absorption (the blend by stellar CaII is included in the model atmosphere calculations).

\begin{figure}
\begin{center}
 \includegraphics[scale=0.35,angle=90]{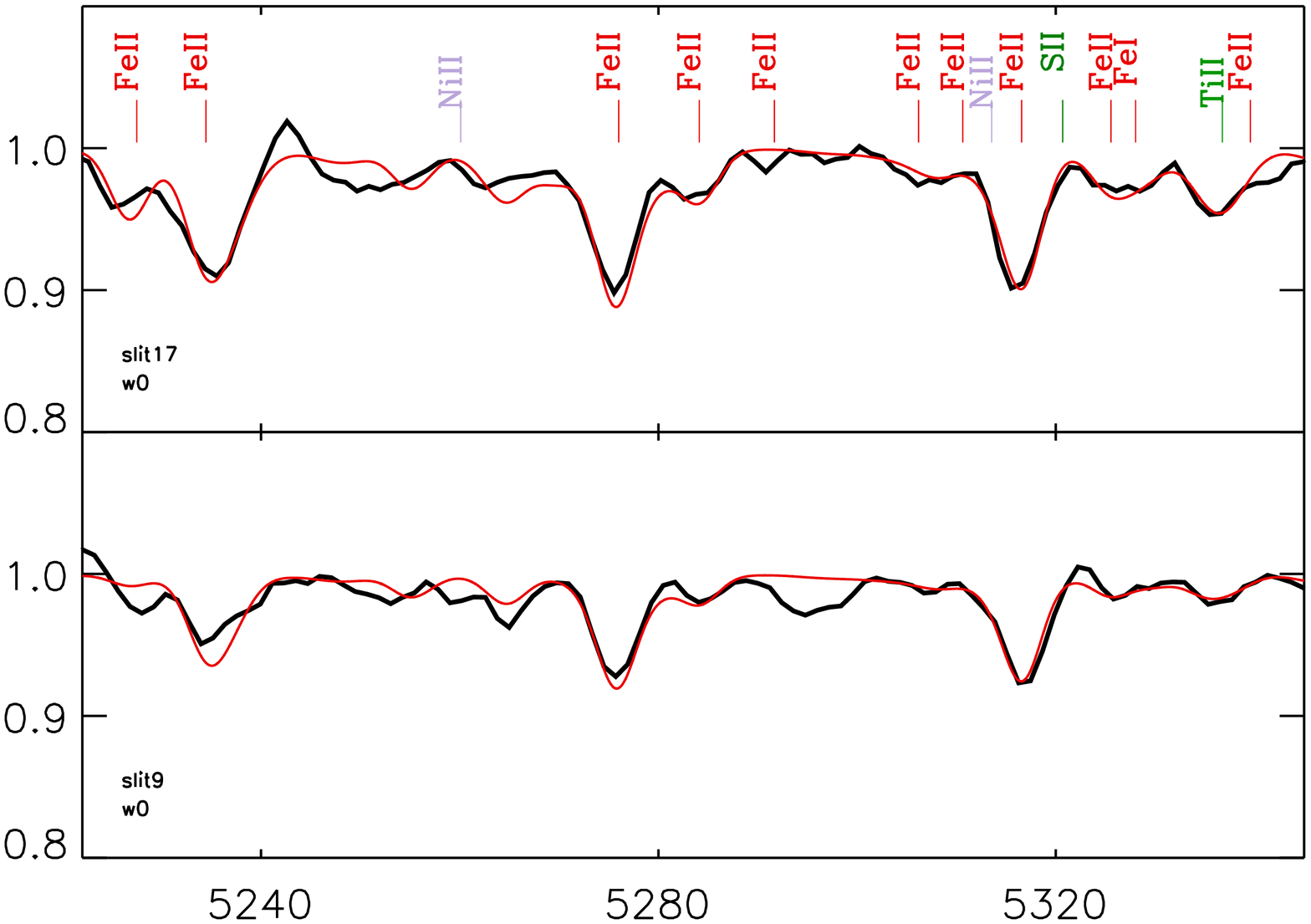}
 \includegraphics[scale=0.35,angle=90]{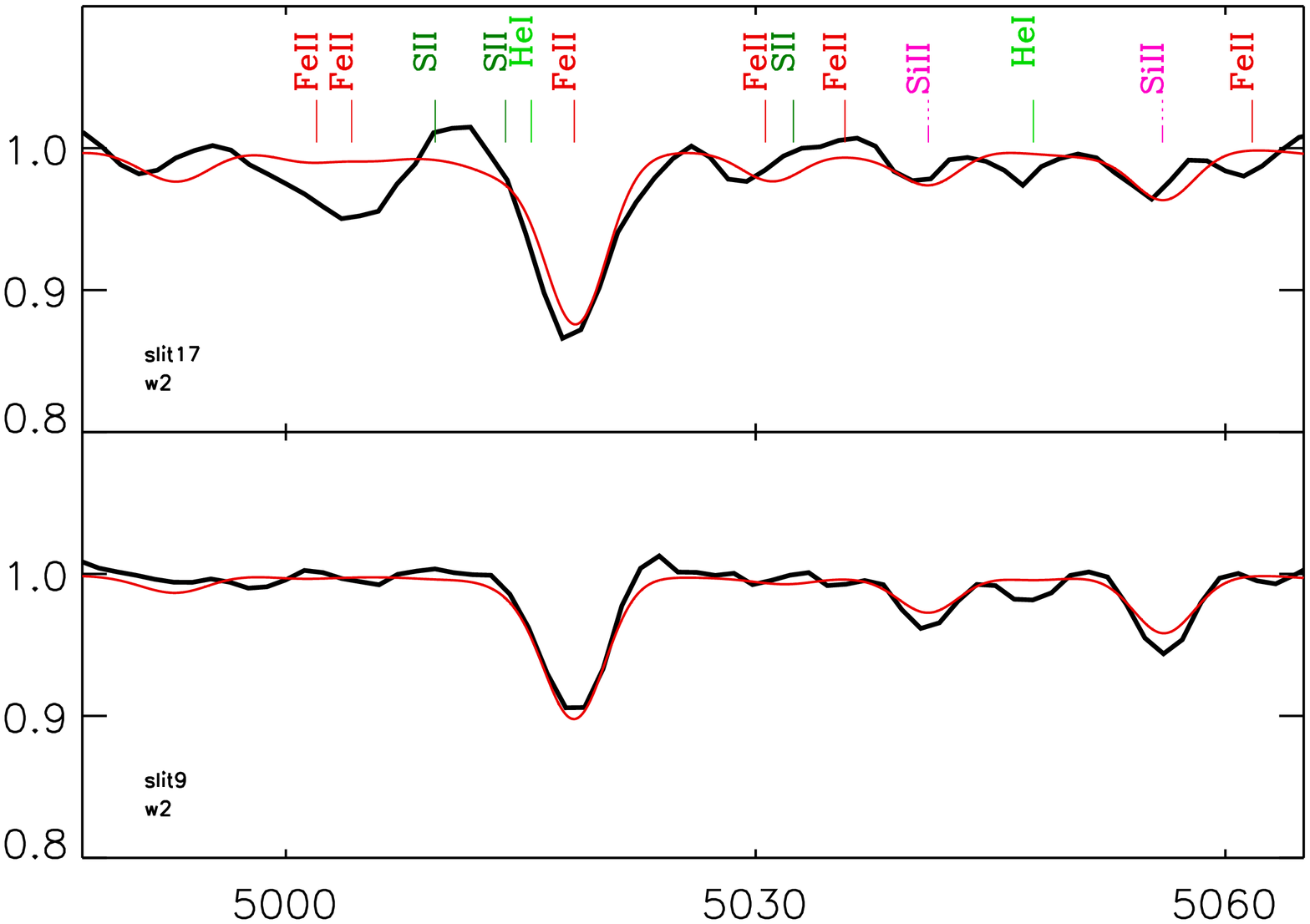}
\caption{Observed metal line spectra (black) of targets Slit 17 and 9 in spectral windows 0 and 2 compared 
with the model calculations (red) obtained for the final stellar parameters of Table 3. \label{metalfit_1}}
 \end{center}
\end{figure}

The strengths of the Balmer lines does not only depend on gravity through the effect of pressure broadening.
It also depends on temperature through the excitation of the first excited level of the hydrogen atom. As the result, 
fits of equal quality as shown in Figure~\ref{balmfit_s9_s17} can be obtained at different values of \teff~for different 
gravity log g. At lower \teff~the log g fit-values are lower, while they are higher at higher \teff. This 
defines a fit curve in the (log g, T$_{\rm eff}$)-plane along which the calculated Balmer lines agree with the observations.
An example is shown in Figure~\ref{balmiso_s9} for Slit 9. The uncertainty of this fit curve is given by the 0.05 dex 
uncertainty of the log g fit at a fixed temeprature. We note that the fit of the Balmer lines is practically independent 
of assumptions on stellar metallicity. Thus, the fit of the Balmer lines provides already a strong restriction of the 
possible values for \teff~and log g. Figure~\ref{balmfit_s14} and \ref{balmfit_smulti} show examples of Balmer line fits 
for the remaining targets.

\begin{figure}
\begin{center}
 \includegraphics[scale=0.35,angle=90]{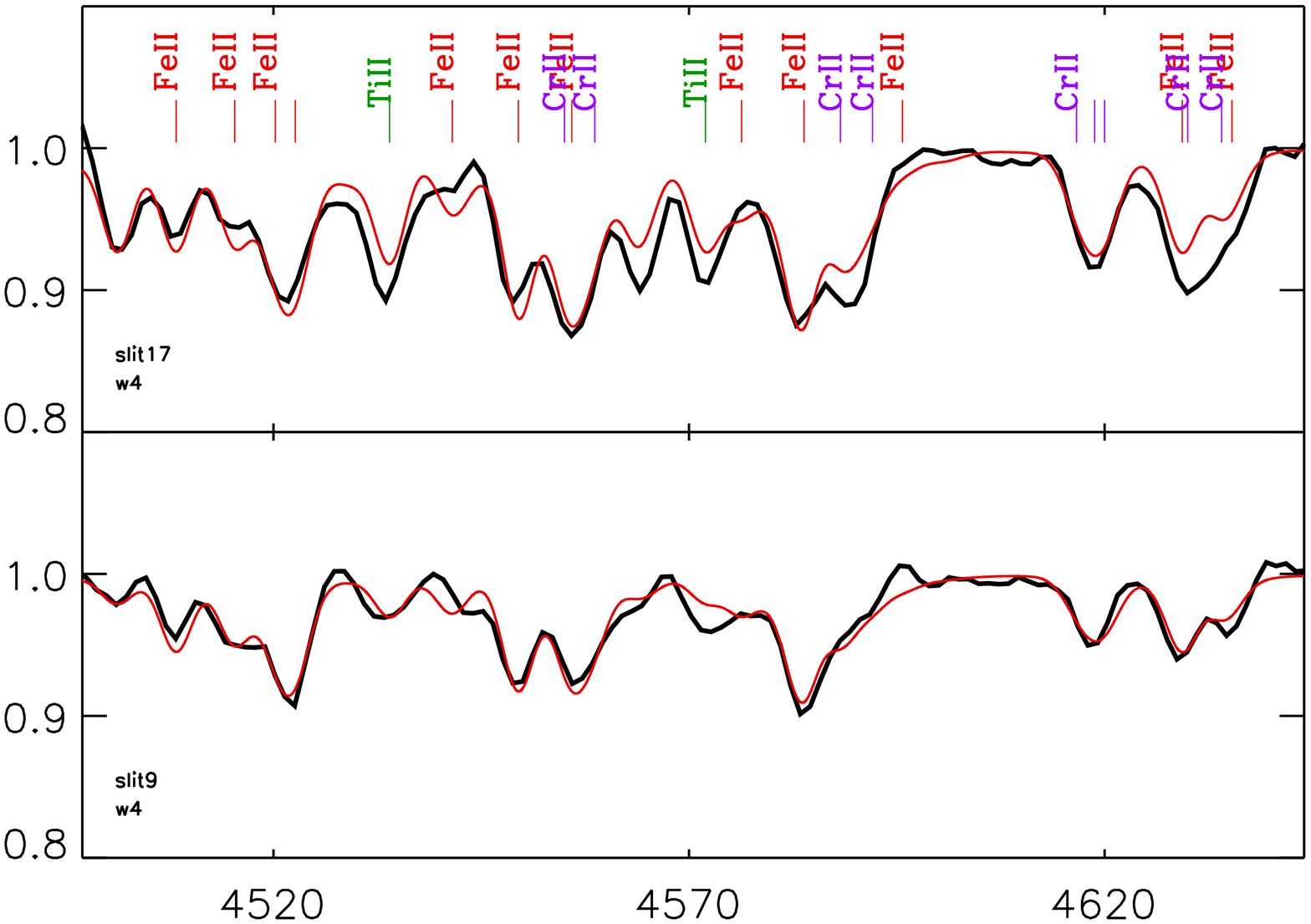}
 \includegraphics[scale=0.35,angle=90]{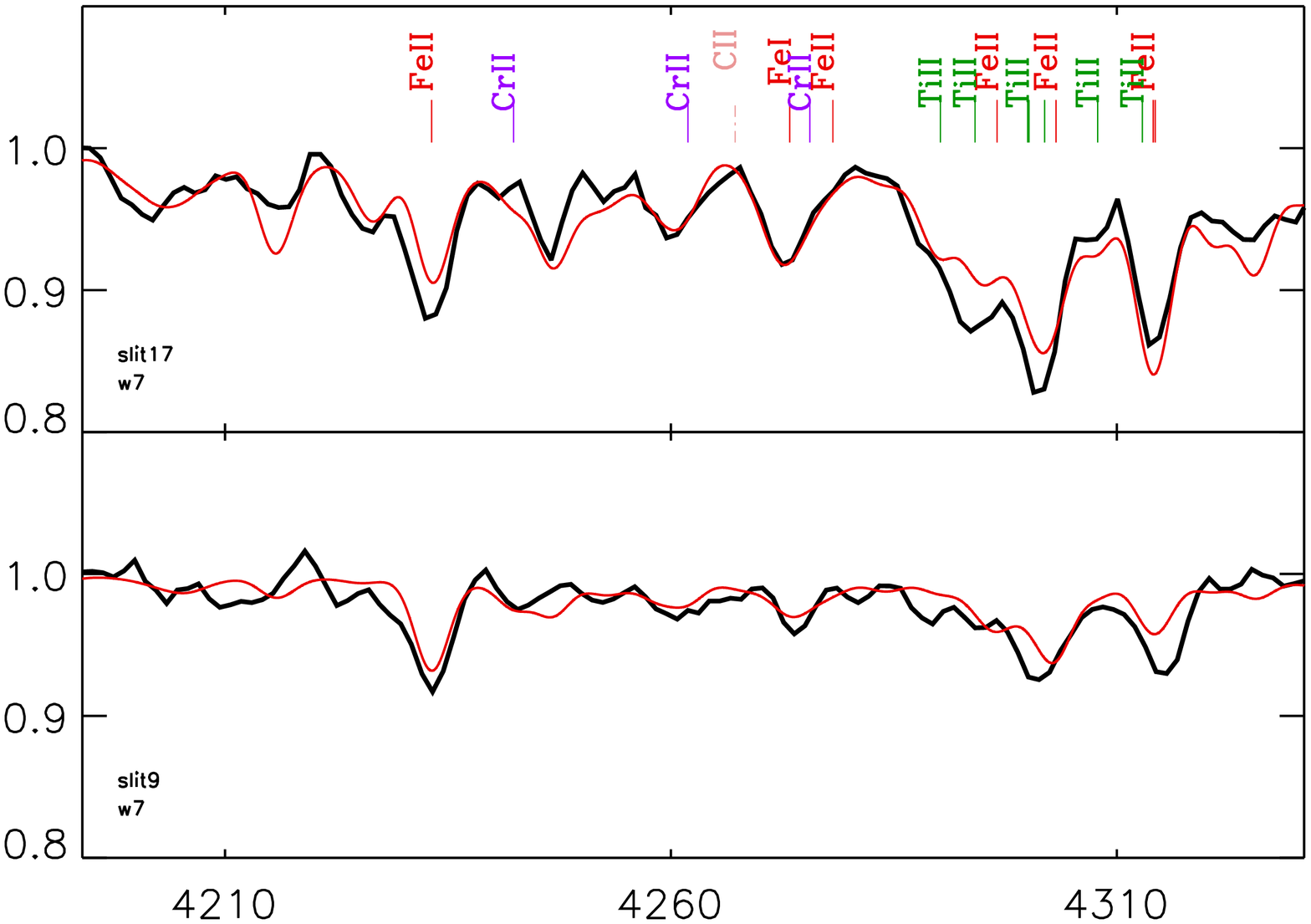}
\caption{Same as Figure~\ref{metalfit_1} but for spectral windows 4 and 7, respectively. \label{metalfit_2}}
 \end{center}
\end{figure}

\begin{figure}
\begin{center}
 \includegraphics[scale=0.35,angle=90]{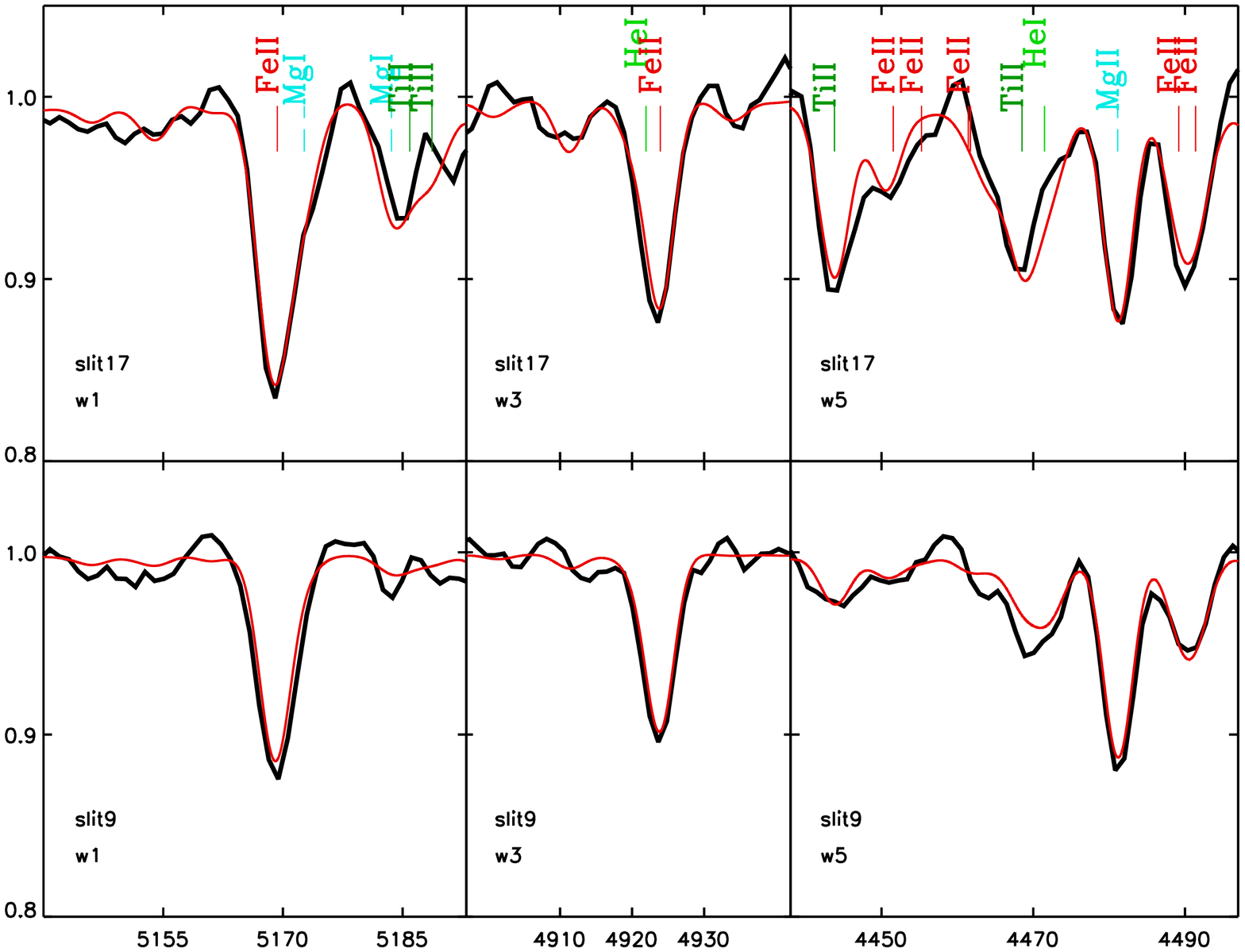}
 \includegraphics[scale=0.35,angle=90]{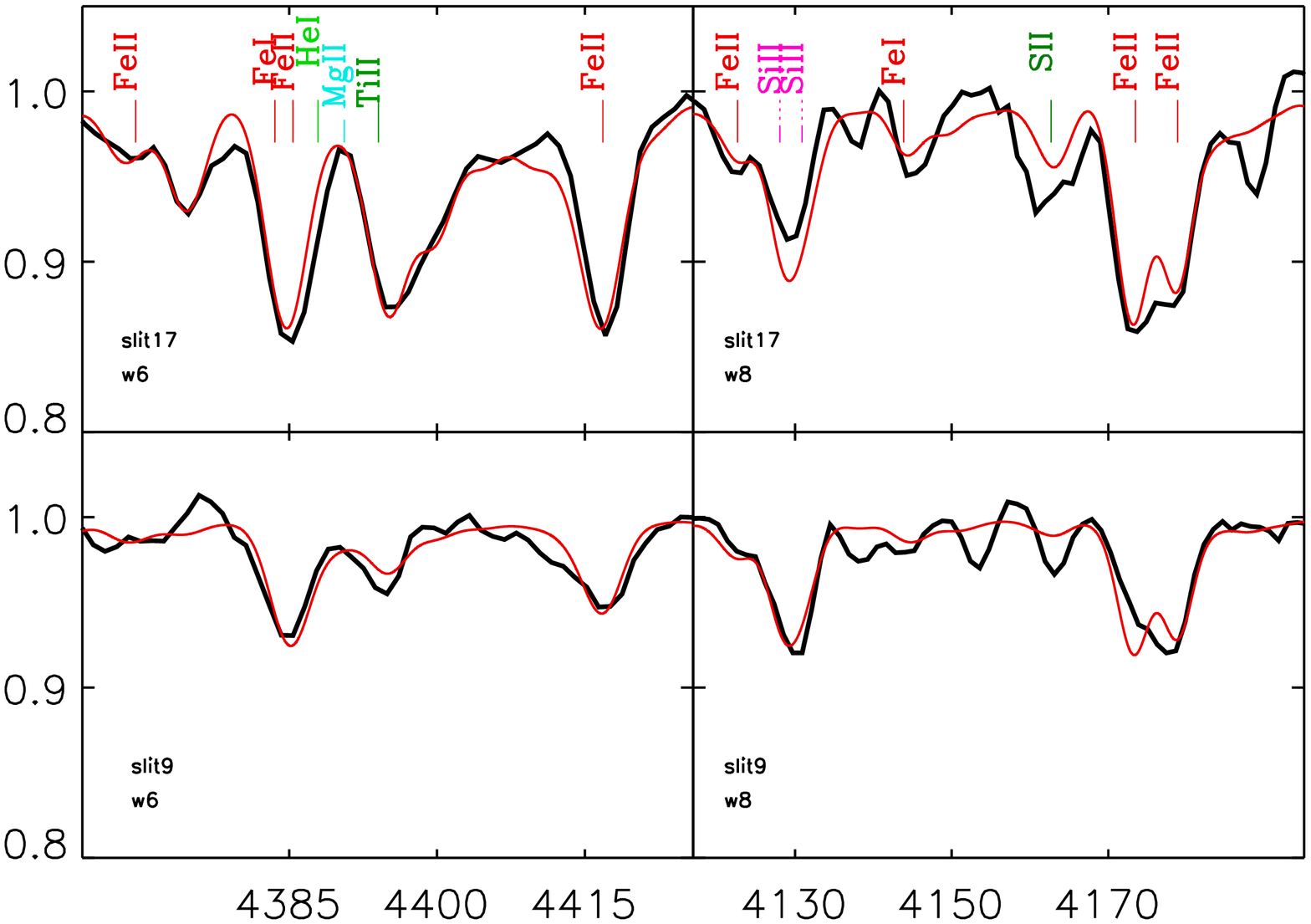}
\caption{Same as Figure~\ref{metalfit_1} but for spectral windows 1, 3, 5 and 6, 8, respectively. \label{metalfit_3}}
 \end{center}
\end{figure}

In the next step, we use the spectrum of the observed metal lines to determine effective temperature and 
metallicity. As in our previous work, we define metallicity as [Z]=log Z/Z$_{\odot}$, where Z$_{\odot}$ is the 
solar metallicity. We then move along the Balmer line fit curve in the (log g, T$_{\rm eff}$)-plane as, for instance, 
in Figure~\ref{balmiso_s9} and compare at each \teff~the observed spectrum 
with synthetic spectra calculated for different [Z] ranging from [Z] = -1.30 to 0.5. For the comparison, we define nine
spectral windows with metal and helium lines (see Figures~\ref{metalfit_1}, \ref{metalfit_2} and \ref{metalfit_3}).  
We split the observed spectrum into different spectral 
windows to allow for a piecewise accurate continuum normalization. The choice of the windows also avoids the Balmer 
lines and nebular emission lines. Sometimes we encounter flaws in some of the observed spectral windows. In such cases,
the window is not used for the analysis. For each temperature 
\teff~and metallicity [Z] we then calculate a $\chi^{2}$-value

\begin{equation}
 \chi^{2}([Z],T_{\rm eff}) = (S/N)^{2} \sum^{n_{pix}}_{j=1} (F_{j}^{obs} - F([Z],T_{\rm eff})_{j}^{calc})^{2}
\end{equation}

\noindent{where S/N is the average target signal-to-noise ratio as given in Table 1. The sum is extended over all 
spectral windows and n$_{pix}$ is the sum of all wavelength points in all spectral windows. 
The $\chi^{2}([Z],T_{\rm eff})$-table obtained in this way allows us to determine the minimum $\chi^{2}_{min}$ in 
the ([Z], T$_{\rm eff}$)-plane and to calculate $\Delta \chi^{2}$ isocontours around this minimum. In order to assess 
which isocontour corresponds to a 1-$\sigma$ uncertainty in this two-dimensional $\chi^{2}$ minimalization process 
we carry out extensive Monte Carlo calculations to simmulate the fitting process. For each target we simulate 1000 
observed spectra by superimposing model spectra with Gaussian Monte Carlo noise corresponding to the observed 
S/N-ratio of our targets. For each simulated spectrum we calculate  $\chi^{2}([Z],T_{\rm eff})$-tables exactly 
in the same way as for the observed spectrum and determine minima $\chi^{2}_{min}$ in the ([Z], T$_{\rm eff}$)-plane.
We compare the distribution of these minima in the ([Z], T$_{\rm eff}$)-plane with isocontours calculated from the 
average of the $\chi^{2}$-tables over all 1000 simulations to determine  which $\Delta \chi^{2}$ isocontour 
encloses 68\% of the MC solutions obtained ($\Delta \chi^{2}$ is the increment relative to the  $\chi^{2}_{min}$ 
value found for the averaged $\chi^{2}$-table). For all targets we find that $\Delta \chi^{2}$ = 3 is a
conservative estimate in reasonable agreement with statistical theory \citep{press07}. More details describing the whole analysis
process can be found in \citet{hosek14}.}

Figure~\ref{iso_s9} shows the isocontours obtained in this way for target Slit 9 indicating an effective temperature of 
\teff=8750$^{+150}_{-250}$K and a metallicity [Z]=-0.30$\pm$0.10. The gravity at this central fit point is log g = 1.00$\pm$0.05 
(see again Figure~\ref{balmiso_s9} as a reference to the log g fit of the Balmer lines). Figure~\ref{iso_multi} displays similar 
isocontour fits for targets Slit 1, 14, 16 and 17 providing well constrained values for \teff, [Z] and gravity log g as 
summarized in Table 2. For illustration of the final metal line fits, Figures~\ref{metalfit_1}, \ref{metalfit_2} and \ref{metalfit_3} show the
comparison of the observed spectra of Slit 9 and 17 in all nine spectral windows with models calculated for final stellar 
parameters given in Table 2. 

Some of the spectral windows selected also contain a contribution of \hei~lines, thus, the metallicity and temperature 
obtained depends also on the helium abundance adopted (we use helium abundances as generally found in high resolution, 
high signal-to-noise-studies of BSGs, see Hosek et al., 2014). However, for targets Slit 1, 9, 14, and 17 the helium lines are weak and 
have no influence on the final result. For Slit 16 there is a weak dependence on helium abundance, but the effect is small compared 
to the uncertainties encountered for this target. Target Slit 3 has an earlier spectral type and the \hei~lines are the strongest lines.
As a consequence, at the S/N of this target metallicity is not well constrained and we can only use the \hei~lines for a relatively uncertain
estimate of effective temperature. We will use this target only for the distance determination using the flux weighted gravity 
log g$_{F}$ = log g - 4log \teff/10$^{4}$K  (see sections 4 and 7). Luckily, the large uncertainty in \teff~does not affect the precision 
with which log g$_{F}$ can be determined, because for effective temperatures larger than 10000K the strength of the Balmer lines 
is constant with constant log g$_{F}$ for a large range of \teff~(see \citep{kud08}, section 6.1 for a physical explanation of this effect). 
This means that even while one needs higher (lower) gravities at higher (lower) temperatures than in in Figure~\ref{balmfit_smulti} to 
fit the Balmer lines of Slit 3, log g$_{F}$ will be the same and is determined accurate to 0.05 dex. Thus, while not useful as a 
metallicity indicator Slit 3 can still be used for the distance determination.


\section{Reddening and Stellar Properties}

\begin{figure}
\begin{center}
 \includegraphics[scale=0.35,angle=90]{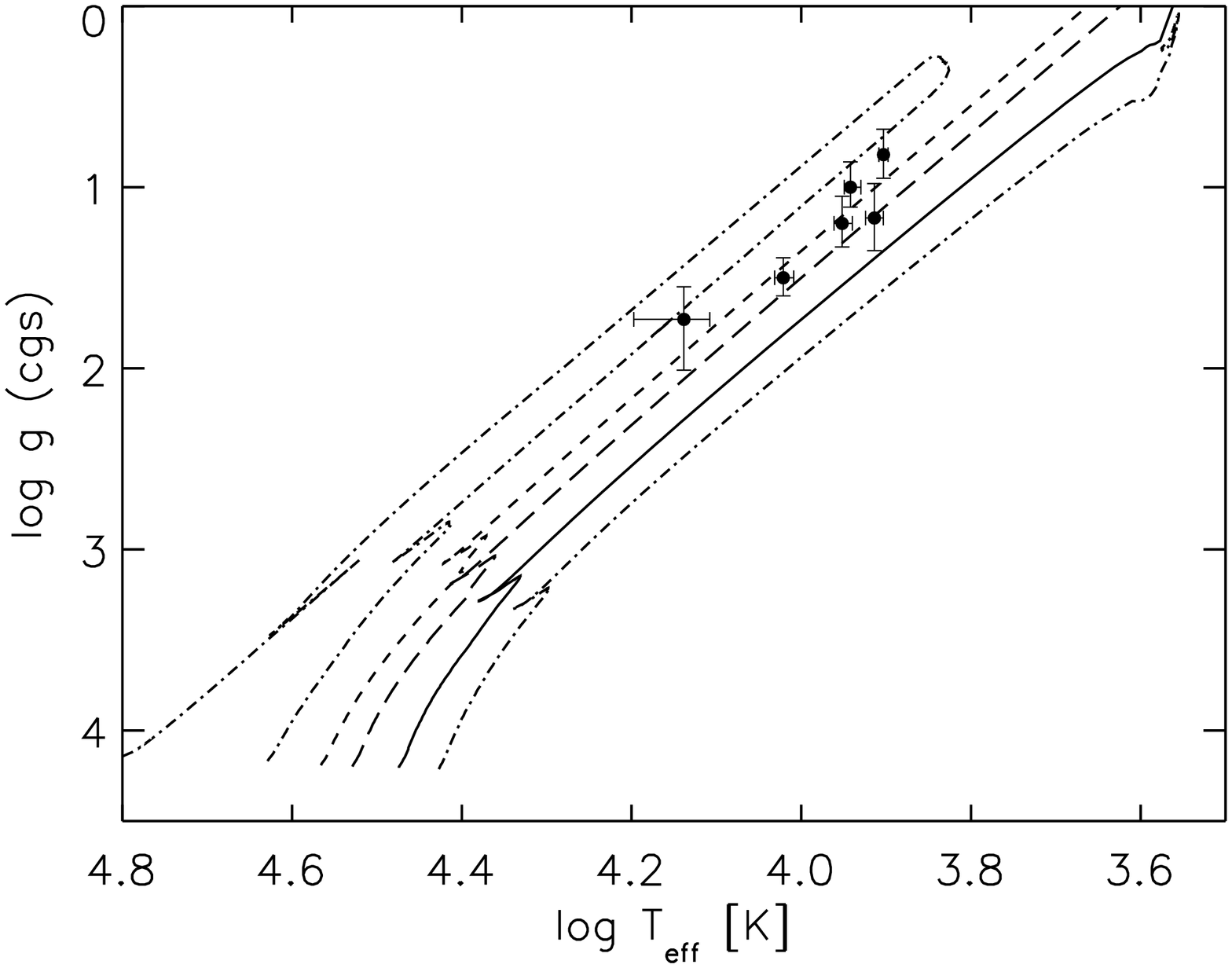}
 \includegraphics[scale=0.35,angle=90]{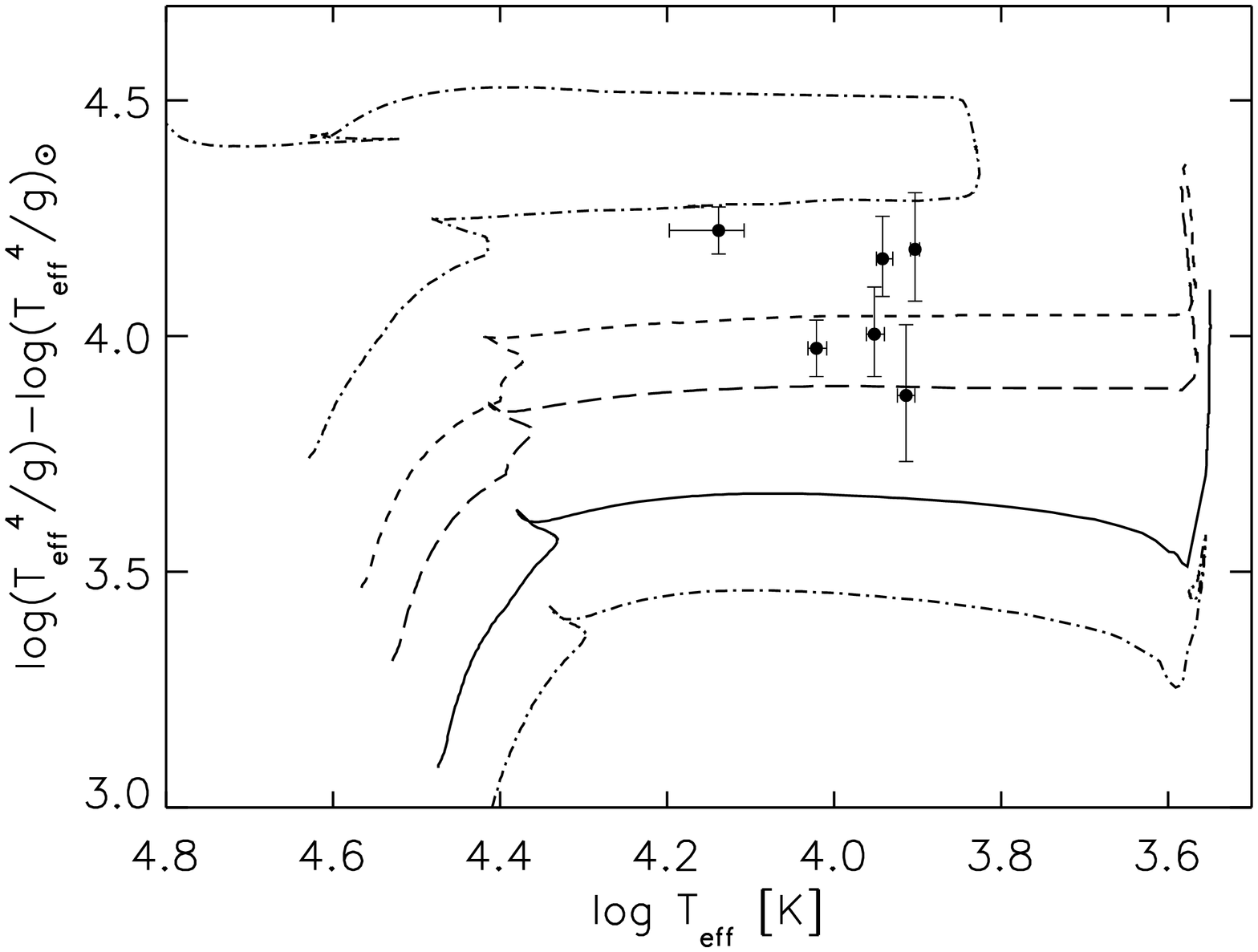}
 \includegraphics[scale=0.35,angle=90]{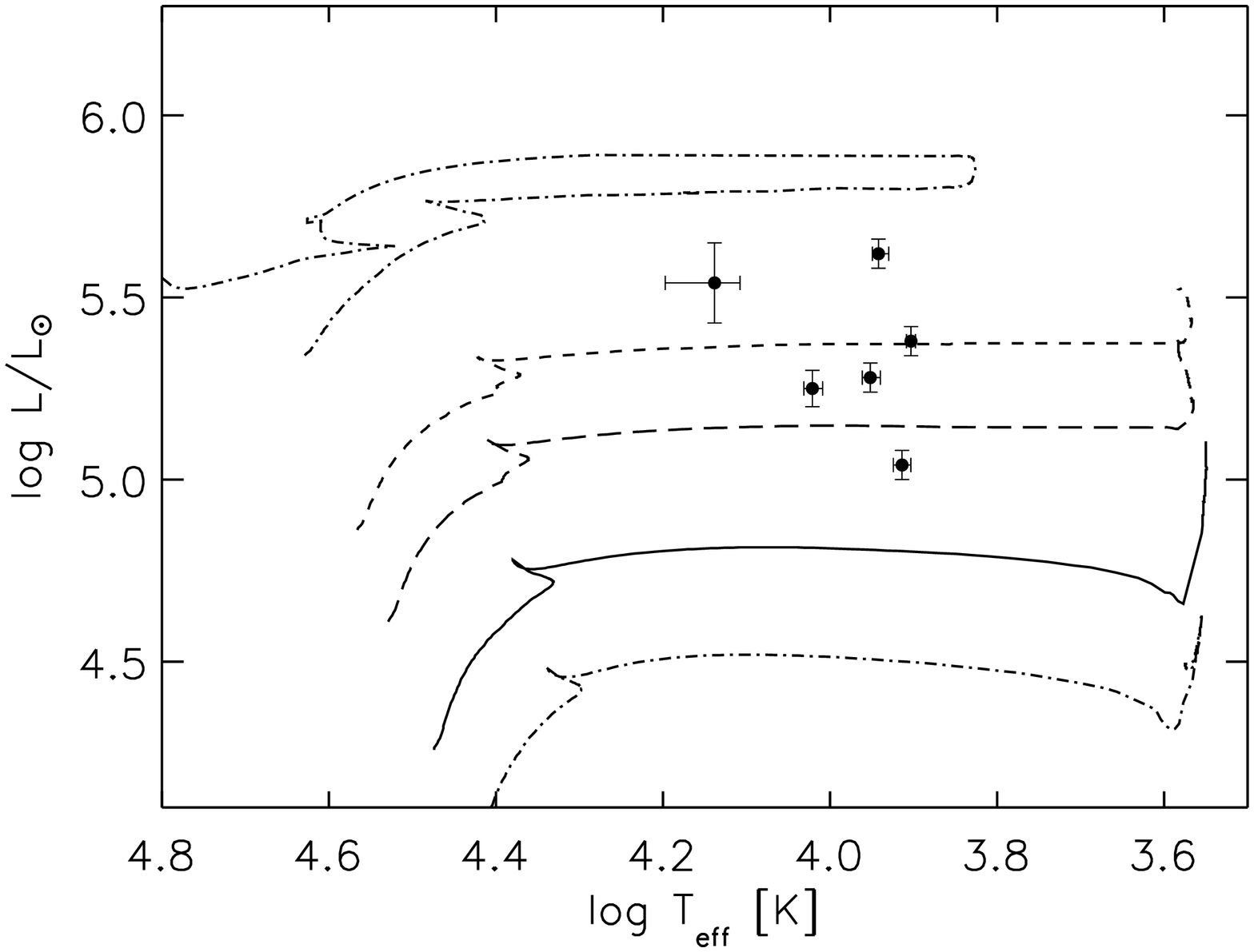}
\caption{Stellar parameters of the observed NGC\,3621 BSGs compared with evolution tracks (with 12, 15, 20, 25, 40 \msun,~respectively) including the effects of rotational mixing (\citealt{meynet05}).
Top: (\teff, log g)-diagram. Middle: sHRD diagram. Bottom: HRD diagram. \label{lgt}}
 \end{center}
\end{figure}

With stellar effective temperatures and gravities obtained from the spectroscopic analysis and summarized in Table 2 we can discuss the stellar 
properties and evolutionary status of the BSGs observed. Figure~\ref{lgt} (top) shows the location of the target stars in the (\teff, log g)-diagram 
compared with evolutionary tracks. The  advantage of (\teff, log g)-diagram is that it is independent of assumption on the distances. This 
allows to investigate the properties of the stellar objects studied without being affected by distance uncertianties. The NGC\,3621 BSG targets 
form an evolutionary sequence of stars, which originally had masses about 25 \msun~and which have now evolved away from the main sequence. Three 
of the objects seem to be somewhat more massive than 25 \msun~although the uncertainties in log g do not allow for a clear conclusion. A 
distance independent alternative to the (\teff, log g)-diagram is the spectroscopic Hertzsprung-Russell diagram (sHRD), which has been recently 
introduced and discussed by \citet{langer14}. This diagram shows the inverse of the flux weighted gravity (as defined in section 3) versus 
effective temperature and is morphologically very similar to the classical HRD. The advantage of this way to study stellar evolution is that 
flux weighted gravities can usually be determined with higher precision than gravities (see Kudritzki et al., 2008, section 6, for a detailed 
discussion). Figure~\ref{lgt} (middle) displays this diagram with the NGC\,3621 BSGs compared to evolutionary tracks. We see a clear indication that 
at least one of the targets is more massive than 25 \msun.

A direct determination of stellar masses, radii and luminosities requires the knowledge of reddening, extinction and distance. With the stellar 
atmospheric parameters \teff, log g and [Z] of Table 2 we calculate model atmospheres colors B-V and V-I. Comparison with the observed colors of 
Table 1 yields reddening E(B-V) directly from B-V, but also from V-I assuming E(B-V)=0.78E(V-I). The average of these two values is given in Table 2. 
The uncertainty is 0.03 mag in agreement with the estimated uncertainty of the HST photometry. Assuming R$_{V}$=3.2 for the ratio extinction  
A$_{V}$ to reddening E(B-V) we can then calculated de-reddened magnitudes. Adding the bolometric correction obtained from the model atmosphere 
calculation gives the de-reddened apparent bolometric magnitudes of Table 2.

\input{tab2}

The average reddening found for the six targets is $\langle$E(B-V)$\rangle$ = 0.18 mag with only a small dispersion of 0.02 mag. This is larger 
than the foreground reddening of 0.07 or 0.08 mag measured by \citet{schlafly11} and \citet{schlegel98}, respectively, but smaller than the value 
of 0.28 mag found by \citet{freedman01} in the HST key project study of Cepheids. \citet{bresolin12} using the \hii~region Balmer decrements 
detected a large range of extinction between 0.0 to 0.5 mag depending on the location within the galaxy. The BSGs of our study (see 
BKMP, Figure 1) are not located in areas where \citet{bresolin12} observed larger extinction (see their Figure 3).
They were selected because of their relatively blue colors and brightness, which may have introduced a bias towards lower reddening. We, thus, 
conclude that our extinction values are not in disagreement with their results. Most importantly, we find that targets Slit 9, 17, and 16 at 
R/R$_{25}$ = 1.29, 0.98 and 0.91, respectively, have reddening values of 0.17 mag. This confirms the conclusion by \citet{bresolin12} that 
the outer disk of NGC\,3621 is not dust free. 

\begin{figure}
\begin{center}
 \includegraphics[scale=0.35,angle=90]{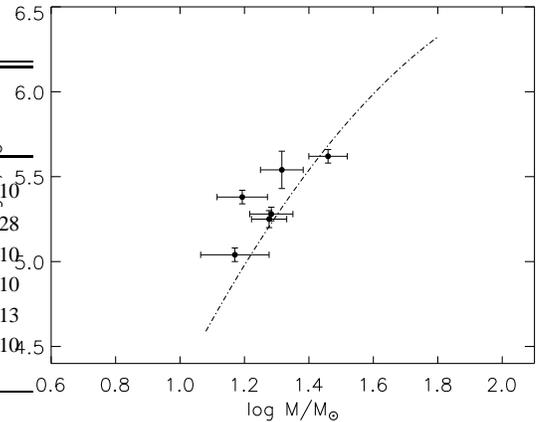}
 \includegraphics[scale=0.35,angle=90]{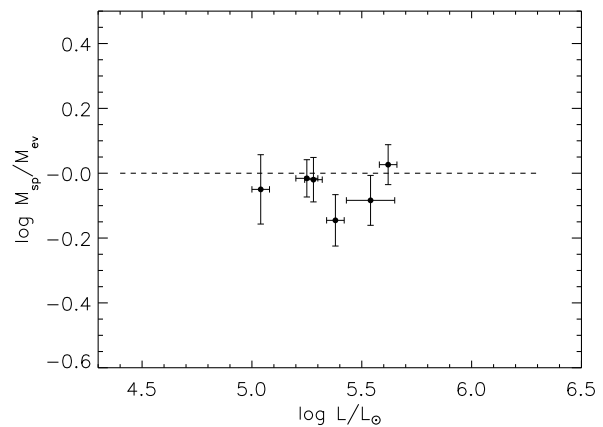}
\caption{Discussion of BSG stellar masses in NGC\,3621. Top: Observed mass-luminosity relationship compared 
with stellar evolution theory using the tracks of Figure 11. Bottom: Logarithm of the ratio of spectroscopic to evolutionary 
masses as a function of luminosity.\label{masslum}}
 \end{center}
\end{figure}

In section 7, we will use the flux-weighted gravity - luminosity relationship (FGLR) to determine a distance modulus to NGC\,3621 
of $\mu$ = 29.07 $\pm{0.1}$ mag. With this distance modulus we obtain the absolute magnitudes and luminosities given in Table 3. The bottom of
Figure~\ref{lgt} then shows the classical Hertzsprung-Russell diagram (HRD) of the BSGs in NGC\,3621 compared with evolutionary tracks. Generally, 
including the additional information obtained from the distance does not change the conclusion about evolutionary status and stellar masses 
except for the target with the lowest \teff, Slit 17. Here the location in HRD indicates a lower mass than one would derive from sHRD. 
We will investigate this further by looking at the directly determined stellar masses. 

The are two ways to determine stellar masses. We can derive stellar radii from the luminosities and effective temperatures and then use 
the gravities to calculate spectroscopic masses. Alternatively, we can use the stellar luminosities and compare with 
the luminosities and the actual masses of the evolutionary tracks at the BSG temperatures to derive evolutionary masses (see \citealt{kud08} for details). 
Both mass estimates are given in Table 3. Because of the relatively large uncertainties of stellar gravities compared with the uncertainties 
of effective temperature and photometry which affect the stellar luminosity, the uncertainties of the spectroscopic masses are larger than those 
of the evolutionary masses. However, it is important to note that the latter do not account for possible systematic uncertainties of the 
evolutionary tracks caused, for instance, by the approximate treatment of rotationally induced mixing or the effects of mass-loss.

\input{tab3}

In Figure~\ref{masslum} we compare the NGC\,3621 BSG luminosities and spectroscopic masses with the mass-luminosity relationship of stellar evolution 
theory. On average, there is agreement between theory and observations with an indication that most of the objects are slightly overluminous similar 
to what has been found in previous studies. 
The most extreme case is Slit 17 as already indicated by the comparison of the sHRD and HRD. This conclusion is confirmed by the plot of the 
ratio of spectroscopic to evolutionary masses also shown in Figure~\ref{masslum}. On average spectroscopy masses are lower by 0.05 dex. This is a 
similar difference as already encountered in the previous extragalactic studies of BSGs (\citealt{kud08}, \citealt{u09}, \citealt{kud12}, \citealt{hosek14}).
It is not clear whether this indicates a systematic deficiency of the spectroscopic log g diagnostics or the evolutionary tracks used to construct 
the evolutionary mass-luminostity relationship. The discrepancy in the case of Slit 17 (0.145 dex or 28\%) seems to be more significant. We note that
Slit 17 with its stellar parameters is at the low temperature, low gravity edge of our model atmosphere grid, where according to \citet{przybilla06} 
pressure inversions are encountered in the model atmosphere structure, which may artificially weaken the Balmer lines in the model spectra. However, 
the pressure inversions in the models used for Slit 17 are only marginal and much smaller than for the \teff~=7700K model discussed by Przybilla et al.
In consequence, no suspicious changes in the strengths of calculated spectral lines as a function of gravity (for fixed effective temperature) are found.
We, thus, conclude that mass-loss processes not included in the evolutionary calculations such as non-conservative binary evolution or evolution back 
to hotter temperature from the red supergiant stage are a more likely reason for the discrepancy.


\section{Metallicity and Chemical Evolution of the Extended Disk of NGC\,3621}

\begin{figure}
\begin{center}
 \includegraphics[scale=0.7]{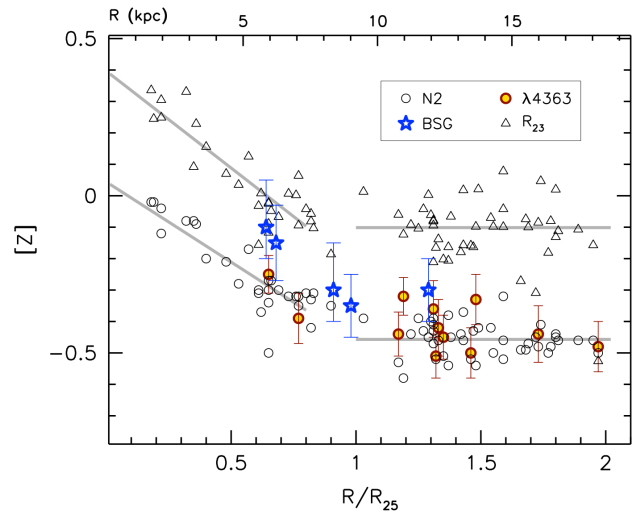}
\caption{Metallicity in NGC\,3621 as a function of galactocentric radius (R/R$_{25}$ on the bottom scale and kpc on the top scale).
Blue supergiant metallicities [Z] = log Z/Z$_{\odot}$ obtained from the spectral analyis described in section 3 are plotted as 
blue stars. Open black circles, triangles and yellow filled red circles represent oxygen abundances as measured by \citet{bresolin12}. 
The transformation of \hii~oxygen abundances N(O)/N(H) to [Z] is given by [Z] = [O/H] - [O/H]$_{\odot}$, 
where [O/H] = 12 + log(N(O)/N(H)) and [O/H]$_{\odot}$ = 8.69 \citep{asplund09}. The values of the open black circles and triangles are 
from the same \hii~regions but use strong emission lines and different calibrations (N2 or R$_{23}$). The values of the yellow filled 
red circles are from the subset of \hii~regions, where the auroral line \oiii~$\lambda$4363 was detected. The grey linear curves are 
regressions for the abundances from the strong line N2 and R$_{23}$ calibrations split into an inner and an outer part. \label{metal}}
 \end{center}
\end{figure}

The \hii~region emission line study by \citet{bresolin12} of the disk of NGC\,3621 found a significant oxygen abundance gradient from the center 
of the galaxy to galacocentric distances of R/R$_{25} \approx 0.8$ corresponding to $\approx$7 kpc. Further out the oxygen abundance profile 
became flat and stayed constant until R/R$_{25} \approx 2.0$ (18 kpc). The result of this work is displayed in  Figure~\ref{metal} and 
shows that the oxygen abundances based on the use of the strong emission lines (strong line methods) depend heavily on the calibration used. 
The N2 calibration gives values about 0.4 dex lower than the R$_{23}$ calibration (see \citealt{bresolin12} for details). A more accurate 
method for \hii~regions is the use of the forbidden \oiii~$\lambda$4363 auroral lines which can be used to constrain the \hii~region electron 
temperatures. To detect and accurately measue the flux of this rather weak emission line requires long exposure times. \citet{bresolin12} 
were able to detect \oiii~$\lambda$4363 in twelve \hii~regions mostly at large galactocentric distances and at low oxygen abundances. This is
very likely
a selection effect since at low metallicities \hii~regions are hotter, which increases the \oiii~$\lambda$4363 emission. The oxygen abundances 
obtained with the help of the auroral $\lambda$4363 line support the N2 calibration which gives lower abundances.

The spectroscopy of the NGC\,3621 BSGs now allows for the first time to compare directly determined stellar metallicities of the young 
stellar population with the \hii~oxygen abundances in the outer disk of this galaxies. This is done in Figure~\ref{metal}. The first conclusion 
is that the BSG metallicities are significantly smaller than those obtained from the R$_{23}$ strong line calibration but larger than those 
obtained from N2. They are also somewhat larger than the more accurate metallicities based on the use of \oiii~$\lambda$4363. The difference is 
small, only $\approx$ 0.1 dex, but appears to be systematic. This could be due to systematic effects influencing the \oiii~$\lambda$4363 method. 
For instance, the depletion of interstellar medium oxygen into interstellar dust could lead to a depletion of ISM oxygen abundance relative to 
stellar metallity. We note that \citet{peimbert10} estimate \hii~region dust depletion factors for oxygen of the order of 0.1 dex, just the same
order of magnitude as encountered here. As another possibility the bias of \oiii~$\lambda$4363 detection towards lower metallicity 
as mentioned above could affect the comparison (see also \citealt{zurita12} for a discussion). Of course, also the BSG metallicity determination 
could be systematically affected, although the high 
resolution, high S/N studies by \citet{przybilla06} indicate that such systematic effects are probably small. It is also important to note that 
the BSG metallicities are based on a fit of the entire metal line spectrum (see Figures~\ref{metalfit_1}, \ref{metalfit_2} and \ref{metalfit_3}) 
and are, thus, mostly representing the abundances of iron and to a weaker extend those of chromium and some higher $\alpha$-elements such as 
magnesium, silicon and titanium. In this sense a deviation of the ratio of oxyygen to iron abundunce caused by a chemical evolution history in
the extended disk of NGC\,3621 different from the solar neighborhood can also not be ruled out. Recent work on Local Group dwarf galaxies 
(\citealt{tolstoy09}, \citealt{hosek14}) indicates that such deviations might exist.  

Independent of the small discrepancy encoutered the quantitative spectroscopy of BSGs as a new and independent tool to investigate the 
metallicity of the young stellar population in NGC\,3621 clearly strengthens the conclusions by \citet{bresolin12}. The BSG metallicities 
confirm the transition from a metallicity gradient in the inner disk to constant metallicity in the outer disk. With [Z] $\approx$ -0.3 obtained 
for the BSGs the metallicity in the outer extended disk of NGC\,3621 is much higher than expected. \citet{bigiel10} have have studied the radial 
profiles of star formation rate and neutral hydrogen gas column density in a larger sample of spiral galaxies and concluded that star formation 
in the outer disks is proceeding very inefficiently. In the case of of NGC\,3621 they measured $\Sigma_{SFR}$ = 10$^{-10}$ \msun~yr$^{-1}$ pc$^{-2}$ for 
the star formation rate column density and $\Sigma_{HI}$ = 5\msun~pc$^{-2}$ for the \hi~ column density at a galactocentric distance of R/R$_{25}$ = 1.0.
Assuming that the outer disks have been originally formed out of very metal poor pristine gas we use a simple 1-zone (``closed box'') chemical 
evolution model (see , for instance, \citealt{pagel09}, page 251) and relate the metallicity mass fraction $Z_{mass}(t)$ of the ISM and the young 
stellar population at time t after the formation of the outer extended disk to $M_{stars}(t)$ and $M_{gas}(t)$, the masses confined in 
stars and ISM gas, respectively

\begin{equation}
Z_{mass}(t) = y_{Z} ln(1 + M_{stars(t)}/M_{gas}(t)).
\end{equation}

\noindent{$y_{Z}$ is the metallicity yield, i.e. the fraction of metals returned to the ISM per newly formed stellar mass. 
We adopt a value of $y_{Z}$ equal to $Z_{mass,\odot}$ = 0.014, the metallicity mass fraction of the sun (\citealt{asplund09}, estimates of 
metallicity yields are discussed by \citealt{maeder92}, \citealt{zoccali03} and \citealt{pagel09})}.
Under the additional simplifying assumption that the presently observed rate is representative for the star 
formation process since the outer disk has formed and that the number ratio of helium to hydrogen in the ISM is 0.1 we can approximate

\begin{equation}
M_{stars(t)}/M_{gas}(t) = \Sigma_{SFR}t/(1.4\Sigma_{HI}) = \frac{1}{70} t_{Gyr}
\end{equation}
\noindent{with $t_{Gyr}$ the age of the outer disk in Gyr.}

Eq. (1) then allows us to calculate the time required to build up the presently observed metallicity of [Z] $\approx$ -0.3 corresponding 
to $Z_{mass}$ = 0.5$Z_{mass,\odot}$ through in situ chemical evolution with the presently observed rate of star formation. We obtain 45 Gyr, much larger 
than Hubble time. (We note that this value is larger than the 10 Gyr estimated by \citet{bresolin12}. The difference is caused by the use of the ``closed box''
relationship of eq. (2) rather than eq. (1) of Bresolin et al., the 0.1 dex higher metallicity, the use of 1.4M$_{HI}$ to represent $M_{gas}$ and 
a metallicity yield reflecting the lower solar metallicity of \citealt{asplund09}). As an estimate of the uncertainties, by doubling the
star formation rate and simultaneously reducing [Z] to -0.4 we obtain 17 Gyr still larger than Hubble time.

\citet{bresolin12} summarize the compelling evidence from cosmological simulations and observations that the 
outer disks of spiral galaxies have ages around 5 Gyr. For such an age eq. (2) predicts a metallicity of [Z] = -1.1 much lower than observed. 
This confirms their conclusion that in situ star formation is unlikely to having produced the metallicity of the young stellar population in the 
outer extended disk and that additional mechanisms are needed. Ejection of chemical enriched 
gas from central galactic regions and subsequent accretion onto the outer disk is discussed by \citet{bresolin12} as the most promising 
mechanism. 

\section{Distance}

The spectroscopically determined temperatures and gravities of the BSGs together with the individual reddening corrections of the photometry
can be used to determine a distance to NGC\,3621 using the flux-weighted gravity--luminosity relationship (FGLR). This new distance 
determination method was introduced by \citet{kud03} and \citet{kud08}.It relates the flux-weighted gravity 
($g_F\,\equiv\,g/{T^4}_{\rm eff}, T_{\rm eff}$ in units of 10$^{4}$K) of BSGs to their absolute bolometric magnitude M$_{bol}$

\begin{equation}
 M_{\rm bol}\,=\,a (\log\,g_F\,-\,1.5)\,+\,b
\end{equation}

\noindent{with $a$ = 3.41$\pm$0.16 and $b$ = -8.02$\pm$0.04 as determined by \citet{kud08}.
The physical background of the FGLR is that massive stars after they leave the main sequence evolve at constant mass and luminosity accross the HRD.
In consequence, because of $g_F \propto M/L$ the flux-weighted gravity remains constant during the horizontal HRD evolution and is independent of temperature.
On the other hand, the luminosity increases strongly with stellar mass and, therefore, $g_F$ decreases with luminosity. This 
establishes the FGLR. So far, distance determinations using the FGLR have already been carried for a number of galaxies 
(see \citealt{kud12} and \citealt{hosek14} and references, therein).}

To determine the distance to NGC\,3621 we use apparent the bolometric magnitudes m$_{\rm bol}$ and flux-weighted gravities log g$_{F}$
and fit a regression of the form

\begin{equation}
 m_{\rm bol}\,=\,a (\log\,g_F\,-\,1.5)\,+\,b_{NGC3621}.
\end{equation}

\begin{figure}
\begin{center}
 \includegraphics[scale=0.35,angle=90]{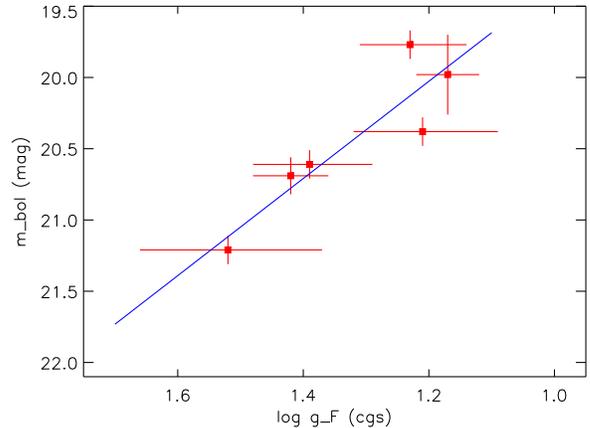}
\caption{Flux weighted Gravity - luminosity relationship (FGLR) of the BSGs observed in NGC\,3621. Plotted is the dereddened apparent 
bolometric magnitude versus flux-weighted gravity of the individual BSGs (red). The blue linear curve is the FGLR of eq. (2) shifted by a 
distance modulus $\mu$ = 29.07 mag. 
\label{fglr}}
 \end{center}
\end{figure}

Since we have only six objects for our regression over a limited range of log g$_{F}$, we adopt the slope of the FGLR of eq. (4) and fit only the 
intercept $b_{NGC3621}$. The result is shown in figure~\ref{fglr}, which indicates that there is a clear relationship between flux-weighted gravity and 
bolometric magnitude well represented by eq. (5). The scatter is $\sigma$ = 0.225 mag. The difference between the fitted intercept $b_{NGC3621}$ and 
the intercept $b$ of the calibration relation eq. (4) yields the distance modulus $\mu$ = 29.07$\pm{0.09}$ mag, where the untertainty is given by $\sigma/\surd 6$.
The uncertainty caused by the errors of a and b (see above) is 0.06 mag.

We compare this result with published Cepheid distances.
During the HST Key Project on the extragalactic distance scale 69 cepheids have been detected in NGC\,3621 \citep{rawson97}. Using their period luminosity relationship 
(PLR) observed in multiple photometric passbands and comparing to the LMC as the distance scale anchor point \citet{freedman01} determined a distance modulus of
$\mu$ = 29.08$\pm{0.06}$ mag. \citet{kanbur03} used the same data set but an improved LMC PLR based on OGLE observations and corrected for charge transfer 
effects (CTE) in the HST camera to obtain a slightly larger distance modulus of 29.15$\pm{0.06}$ mag. Both distance moduli quoted  were obtained 
for the case assuming that the Cepheid PLR does not vary with metallicity, but in both papers a metallicity correction was also applied yielding distance moduli 
0.03 and 0.05 mag larger. We note, however, that the metallicity corrections are very uncertain for a variety of reasons (see Kudritzki et al., 
2012, 2013 for a discussion). \citet{paturel02} used a HIPPARCOS based calibration of Milky Way Cepheids for the PLR and 
obtained $\mu$ = 29.13$\pm{0.06}$ mag. \citet{kanbur03} also included a distance determination based on an alternative Milky way PLR calibration of yielding 
$\mu$ = 29.22$\pm{0.06}$ mag. It seems that within the range of uncertainties the distance obtained with the new BSG FGLR method is in good 
agreement with the HST Cepheid work.

An alternative method to obtain distances is the use of the magnitude of the observed tip of the red giant branch (TRGB). The earlier 
work by \citet{sakai04} and \citet{rizzi07} resulted in relatively large distance moduli, 29.36$\pm{0.11}$ mag and 29.26$\pm{0.12}$ mag, respectively. However, 
the most recent work published in the EDD database (http://edd.ifa.hawaii.edu, see \citealt{tully09} for a description) provides a distance 
$\mu$ = 29.11$\pm^{0.06}_{0.04}$ mag in good agreement with our FGLR result and the Cepheid studies. 

\section{Discussion and Conclusions}

The spectroscopic study presented in work this provides the first direct information about the chemical composition of the young 
stellar population in the disk of the spiral galaxy NGC\,3621. We observe the transition from an inner disk with a 
metallicity gradient to an outer disk with constant metallicity. This confirms the result by \citet{bresolin12} obtained from 
their extensive study of \hii~regions. The BSG in the outer extended disks are reddened by $\sim$ 0.17 mag indicating a significant amount 
of interstellar dust. Their metallicity is only a factor of two lower than the metallicity of the sun. 
This is a factor of six higher than one would expect from in situ chemical 
evolution at the present level of star formation and neutral hydgrogen density over the lifetime of the extended disk of 4-6 Gyr 
predicted by cosmological simulations. To produce such a high metallicity under these conditions would require 45 Gyr much
longer than Hubble time. This requires additional mechanisms to build up the high level of metallicity observed. One possible 
mechanism suggested by cosmological simulations is the the ejection of metal enriched gas from the inner disk through galactic 
winds and subsequent accretion in the outer disk.

Using stellar temperatures, gravities and reddening corrected photometry of the individual the individual BSGs as obtained from the 
quantitative spectroscopic analysis a distance to NGC\,3621 of $D=6.52\pm0.28$\,Mpc is obtained. This distance is in good agreement 
with Cepheid distance determinations based on the sample of cepheids detected by the HST Key Project on the extragalactic distance 
scale. It also agrees with a most recent determination using NGC\,3621 HST color magnitude diagrams and the tip of the red giant branch.


\acknowledgments This work was supported by the National Science Foundation under grant AST-1008798 to RPK and FB.




{\it Facilities:} \facility{ESO VLT (FORS)}, \facility{HST (ACS)}.

\clearpage












\end{document}

%% file: tab1.tex
\begin{deluxetable}{cccccccccc}
\tabletypesize{\scriptsize}
\tablecolumns{9}
\tablewidth{200pt}
\tablecaption{NGC~3621 - Spectroscopic targets}

\tablehead{
\colhead{name$^{\tablenotemark{a}}$}            &
\colhead{$\alpha_{2000}$$^{\tablenotemark{a}}$}           &
\colhead{$\delta_{2000}$$^{\tablenotemark{a}}$}     &
\colhead{R/R$_{25}$$^{\tablenotemark{b}}$}     &
\colhead{sp.t.}             &
\colhead{m$_{V}$}     &
\colhead{B-V} &
\colhead{V-I}    &
\colhead{S/N}   & 
\colhead{}\\
\colhead{}         &
\colhead{ h min  sec}         &
\colhead{ \arcdeg~ \arcmin~ \arcsec }     &
\colhead{}   &
\colhead{}         &
\colhead{mag}         &
\colhead{mag}		&
\colhead{mag}			&
\colhead{}\\[1mm]
\colhead{(1)}	&
\colhead{(2)}	&
\colhead{(3)}	&
\colhead{(4)}	&
\colhead{(5)}	&
\colhead{(6)}	&
\colhead{(7)}	&
\colhead{(8)}	&
\colhead{(9)}}
\startdata
\\[-1mm]
  Slit 1  & 11 18 14.3 & -32 45 40.1 & 0.68 & A1 & 21.236 & 0.240 & 0.282 &  55 \\[2pt]
  Slit 3$^{\tablenotemark{c}}$  & 11 18 10.3 & -32 45 31.4 & 0.63 & B5 & 21.573 & 0.084 & 0.199 &  45 \\[2pt]
  Slit 9  & 11 18 28.7 & -32 48 04.6 & 1.29 & A2 & 20.390 & 0.244 & 0.294 & 125 \\[2pt]
  Slit 14 & 11 18 26.6 & -32 50 05.8 & 0.91 & A4 & 21.859 & 0.267 & 0.326 &  45 \\[2pt]
  Slit 16 & 11 18 23.3 & -32 51 02.7 & 0.64 & B9 & 21.594 & 0.172 & 0.174 &  50 \\[2pt]
  Slit 17 & 11 18 27.5 & -32 51 05.4 & 0.98 & A5 & 20.888 & 0.289 & 0.320 &  80 \\[2pt]
\enddata
\tablenotetext{a}{Bresolin et al. (2012)}
\tablenotetext{b}{Galactocentric distance, in units of R$_{25}$ = 4.89 arcmin $\simeq$ 9.28 kpc (distance modulus 29.07 mag).\\ 
A position angle PA = 345\arcdeg, an inclination i =65\arcdeg and central coordinates $\alpha_{2000}$ = 11h18min16.52sec,\\
 $\delta_{2000}$ = -32\arcdeg 48\arcmin 50.7\arcsec were assumed (Hyperleda data base, Paturel et al., 2003)}
\tablenotetext{c}{no HST photometry, see text}
\end{deluxetable}

%% file: tab2.tex
\begin{deluxetable}{cccccccccc}
\tabletypesize{\scriptsize}
\tablecolumns{8}
\tablewidth{200pt}
\tablecaption{Stellar Parameters}

\tablehead{
\colhead{name}            &
\colhead{\teff}           &
\colhead{log g}     &
\colhead{log g$_{F}$}     &
\colhead{[Z]}             &
\colhead{E(B-V)}     &
\colhead{BC} &
\colhead{m$_{bol}$}\\
\colhead{}         &
\colhead{K}         &
\colhead{cgs}     &
\colhead{cgs}   &
\colhead{dex}         &
\colhead{mag}         &
\colhead{mag}		&
\colhead{mag}\\[1mm]
\colhead{(1)}	&
\colhead{(2)}	&
\colhead{(3)}	&
\colhead{(4)}	&
\colhead{(5)}	&
\colhead{(6)}	&
\colhead{(7)}	&
\colhead{(8)}}
\startdata
\\[-1mm]
 Slit 1  &  8950$\rm^{200}_{250}$    & 1.20$\rm^{0.13}_{0.15}$ & 1.39$\rm^{0.09}_{0.10}$ & -0.15$\rm^{0.12}_{0.12}$ & 0.19 & -0.11 & 20.61$\pm{0.10}$ \\[2pt]  
 Slit 3$^{\tablenotemark{a}}$  & 13750$\rm^{2000}_{1000}$  & 1.73$\rm^{0.28}_{0.18}$ & 1.17$\rm^{0.05}_{0.05}$ &                          & 0.19 & -0.99 & 19.98$\pm{0.28}$ \\[2pt]  
 Slit 9  &  8750$\rm^{150}_{250}$    & 1.00$\rm^{0.11}_{0.14}$ & 1.23$\rm^{0.08}_{0.09}$ & -0.30$\rm^{0.10}_{0.10}$ & 0.17 & -0.09 & 19.77$\pm{0.10}$ \\[2pt]  
 Slit 14 &  8200$\rm^{200}_{200}$    & 1.17$\rm^{0.18}_{0.19}$ & 1.52$\rm^{0.14}_{0.15}$ & -0.30$\rm^{0.15}_{0.10}$ & 0.22 &  0.04 & 21.21$\pm{0.10}$ \\[2pt] 
 Slit 16 & 10500$\rm^{250}_{300}$    & 1.50$\rm^{0.10}_{0.11}$ & 1.42$\rm^{0.06}_{0.06}$ & -0.10$\rm^{0.15}_{0.10}$ & 0.17 & -0.37 & 20.69$\pm{0.13}$ \\[2pt]   
 Slit 17 &  8000$\rm^{100}_{100}$    & 0.82$\rm^{0.13}_{0.14}$ & 1.21$\rm^{0.11}_{0.12}$ & -0.35$\rm^{0.10}_{0.10}$ & 0.18 &  0.06 & 20.34$\pm{0.10}$ \\[2pt]
\enddata
\tablenotetext{a}{\teff ~estimated from \hei lines; no metallicity determined}
\end{deluxetable}

%% file: tab3.tex
\begin{deluxetable}{cccccccc}
\tabletypesize{\scriptsize}
\tablecolumns{8}
\tablewidth{0pt}
\tablecaption{Absolute magnitudes, luminosities, radii and masses}

\tablehead{
\colhead{name}            &
\colhead{M$_{V}$}           &
\colhead{M$_{bol}$}     &
\colhead{log L/L$_\odot$}     &
\colhead{R}             &
\colhead{M$_{spec}$}     &
\colhead{M$_{evol}$}\\
\colhead{}         &
\colhead{mag}         &
\colhead{mag}     &
\colhead{dex}   &
\colhead{R$_{\odot}$}         &
\colhead{M$_{\odot}$}         &
\colhead{M$_{\odot}$}\\[1mm]	
\colhead{(1)}	&
\colhead{(2)}	&
\colhead{(3)}	&
\colhead{(4)}	&
\colhead{(5)}	&
\colhead{(6)}	&
\colhead{(7)}}	
\startdata
\\[-1mm]
 Slit 1  & -8.35 & -8.46 & 5.28$\pm{0.04}$ & 128.8$\pm{9.1}$  & 19.2$\pm{3.2}$ & 20.1$\pm{0.7}$\\[2pt]
 Slit 3  & -8.10 & -9.09 & 5.54$\pm{0.11}$ & 103.7$\pm{27.2}$ & 20.7$\pm{3.4}$ & 25.1$\pm{2.4}$\\[2pt]
 Slit 9  & -9.21 & -9.30 & 5.62$\pm{0.04}$ & 281.6$\pm{18.9}$ & 28.8$\pm{4.2}$ & 27.1$\pm{1.0}$\\[2pt]
 Slit 14 & -7.90 & -7.86 & 5.04$\pm{0.04}$ & 165.0$\pm{11.4}$ & 14.8$\pm{4.1}$ & 16.6$\pm{0.5}$\\[2pt]
 Slit 16 & -8.01 & -8.38 & 5.25$\pm{0.05}$ & 128.3$\pm{10.6}$ & 18.9$\pm{2.5}$ & 19.6$\pm{0.8}$\\[2pt]
 Slit 17 & -8.75 & -8.69 & 5.38$\pm{0.04}$ & 254.7$\pm{13.5}$ & 15.6$\pm{3.1}$ & 21.8$\pm{0.7}$\\[2pt]
\enddata
\end{deluxetable}